\documentclass[twocolumn]{aastex631}
\usepackage{comment}
\usepackage{soul}

\begin{document}

\title{Effects of an Explicit Time-Dependent Radiation Pressure Force on Trajectories of Primary Neutral Hydrogen in the Heliosphere}

\author[0009-0007-0108-2318]{Lucas Dyke}
\affiliation{Department of Physics and Astronomy, Dartmouth College, 6127 Wilder Laboratory, Hanover NH 03755}
\email{lucas.r.dyke.gr@dartmouth.edu}

\author[0000-0001-7364-5377]{Hans-Reinhard M\"{u}ller}
\affiliation{Department of Physics and Astronomy, Dartmouth College, 6127 Wilder Laboratory, Hanover NH 03755}

\begin{abstract}
    Radiation pressure exerted by solar photon output is salient to the motion of primary neutral hydrogen atoms streaming into the inner heliosphere directly from the Local Interstellar Medium (LISM). The action of a time-dependent radiation pressure force, when coupled with the usual gravitational force, changes the characteristic velocities, and therefore energies, of the atoms when they reach regions in which explorer probes are present. A study is presented that uses a 2D code to backtrace neutral hydrogen trajectories from representative target points located 1 au from the Sun. It makes use of both a radiation pressure function and a function for the photoionization rate at 1 au that oscillate with time based on measurements over a typical solar cycle, as well as a time-independent charge exchange ionization rate at 1 au. Assuming a Maxwellian distribution in the distant upwind direction, phase space data is calculated at the target points, at different moments in time. The dependence of the force on the radial particle velocity has been omitted in the analysis, such that the emphasis is on the effects of the global solar UV intensity variations through the solar cycle. This process allows for analysis of direct and indirect Maxwellian components through time and space in the time-dependent force environment. Additionally, pseudo-bound orbits caused by energy losses associated with this force environment are observed and their properties evaluated with the aim of determining their effects on potential measurements by explorer probes.
\end{abstract}


\section{Introduction} \label{sec:intro}

The local interstellar medium (LISM) surrounding the Sun injects interstellar neutral atoms and also neutral particles derived from them into the heliosphere. These neutral atoms form a heliospheric hydrogen wall \citep{baranov1991} and similar remote overdensities in other elemental species (e.g., helium wall, oxygen wall \citep{mueller2004}). They contribute to other measurable effects such as the glow of scattered solar FUV photons \citep{bertaux1971, thomas1971} or the production of pickup ions \citep{moebius1985}. Neutral atom detectors onboard IBEX and IMAP \citep{mccomas2009,mccomas2018} are able to detect these low-energy neutral particles directly and characterize their velocity distribution functions $f(\mathbf{x},\mathbf{v})$ (VDF) through measuring the energy- and direction-dependent particle fluxes. Ulysses also has this capability, using a different measurement strategy \citep{witte1996,wood2015}. These observations are carried out at $\sim$1 au (Ulysses: 1.4 – $\sim$2 au) where neutral atom populations have been altered by their passage through the heliosphere.

Trajectory methods are used successfully to connect LISM neutral helium with helium measurements by IBEX-Lo \citep{bzowski2015,lee2015}. Helium is the prime example of a species that can be treated with conservation methods, as the only force acting on helium is solar gravity. The present paper focuses on the modeling of neutral hydrogen (H), hydrogen being the dominant species in the heliosphere. The main contrast with helium is that through the solar Lyman-$\alpha$ (Ly-$\alpha$) profile, hydrogen atoms experience a combination of gravity and radiation pressure, with the latter introducing time dependence into the force acting on the atom. The time dependence enters in two ways: in the form of the strength of the solar Ly-$\alpha$ line varying on the time scale of the solar cycle \citep{lemaire2002,lemaire2005,lemaire2015}, and through the Doppler effect related to the radial velocity component of the particle on its trajectory. Different radial velocities of the particle scan different parts of the double-peaked, non-flat Ly-$\alpha$ profile, which is depicted in Figure 5 in \citet{lemaire2015}. In the present paper we narrow our focus to the first dependence only, considering a force on H in which time variation only enters through an oscillation of the radiation pressure over the solar cycle but whose underlying Ly-a profile is assumed flat and hence does not produce a dependence on radial velocity.

This directly deviates from other studies that have focused on studying the results of the time dependence of the radiation pressure force due to the Doppler effect. \citet{tarnopolski2007,tarnopolski2009} study the effect of both H and deuterium trajectories subject to a Doppler-dependent radiation force for two distinct, constant solar profiles, representative of solar minimum levels and solar maximum, respectively. Their analysis revealed cases where deuterium experiences a transition from hyperbolic to bound (pseudo-bound) orbits, a phenomenon that we also encounter in the present study even with our different radiation pressure mechanism (see Section \ref{sec:pb}).

The goal of this paper is to characterize the behavior of VDF structures throughout the solar cycle under our simplifying assumption, and create a basis on which to analyze further studies which will include all realistic heliospheric ingredients of radiation pressure. In this future model, radiation pressure will depend on data-driven solar observations and include the non-flat profile. Prior studies \citep{tarnopolski2009,izmodenov2013} have modeled the behavior of interstellar neutral hydrogen with many of these effects included, albeit focusing on bulk flow properties.

A difference between prior studies and this paper is the method used to examine particle trajectories. While models like \citet{izmodenov2013} use kinetic particle methods to analyze neutral ISN H behavior, here we use trajectory methods. This choice avoids the need for averaging over a volume to obtain sufficient particle statistics, and yields as high a resolution in velocity space as is desired through a set of individual trajectories. The physics-based, {\em ab-initio} calculations will complement recent studies that take spacecraft observations like those by IBEX to indirectly infer the radiation pressure force \citep{rahmanifard2019,rahmanifard2023,katushkina2021}.

One of the advantages of the use of trajectory methods is the ability to analyze VDF structures in space and time and assess their relevance for spacecraft measurements. Analysis of VDF structures and how they vary with the solar cycle allows us to pinpoint the structures at varying spatial points at different instants of time and eventually determine whether spacecraft instrumentation can measure particles at these times, in addition to allowing integration up to bulk quantities in future studies. Prior studies \citep{tarnopolski2009,bzowski2015,izmodenov2013} focus on studies of bulk quantities (moment integrals of VDF) and not as much on the VDF themselves.

In the rest of this section, the relevant physics of neutral hydrogen trajectories in the vicinity of the Sun will be discussed. Section \ref{sec:code} discusses the code, and Section \ref{sec:results} presents the results of the simulations and their analysis and discussion.

\subsection{Definition of Coordinate System}

We start by defining our coordinate system. We take the upwind axis to be the $x$-axis, such that the inflow from the interstellar medium (ISM) is directed along the negative $x$ direction toward the Sun, which is located at the origin. Then, we take the axis perpendicular to this flow to be the $y$-axis such that the axis itself lies fully within the ecliptic plane. Thus, any spatial points with a negative $x$-coordinate are considered to be downwind, any with a positive $x$-coordinate are considered upwind, and any with $x=0$ are labelled as crosswind points.

The validity of the 2D formulation hinges on the fact that we are dealing with a purely central force. The radiation pressure force acts as a central force and scales as $1/r^2$, similar to the gravitational force, such that the net force on any neutral hydrogen particle at any point in time is given by
\begin{equation}
    \mathbf{F}_{net} = -\frac{G M_{\odot} m_H}{r^2} (1 - \mu(t)) \hat{r}
    \label{rpforce}
\end{equation}
where $\hat{r}$ is the unit vector pointing radially outward from the Sun, $G M_{\odot}$ is the solar gravity parameter \citep{park2021}, $m_H$ is the mass of a hydrogen atom, and $\mu (t)$ expresses the ratio of the radiation pressure force $F_{rp}$ and the magnitude of the gravitational force $F_g$, $\mu = \frac{F_{rp}}{F_g}$. Since the net force is still a central force, angular momentum is still conserved and orbits will hence be restricted to a 2D plane. Thus, we are allowed to consider movement within a 2D plane aligned with the inflow for the purpose of this code without losing generality. 

\subsection{Direct and Indirect Trajectories}

We consider the effects of our force environment on neutral hydrogen particles specifically directly streaming in from the LISM, which can be referred to as ``primary" neutrals. In the warm model of interstellar neutrals, one considers a Maxwellian distribution at infinity for the velocity of ISM particles, centered around the representative value of $\mathbf{u}_{ISM} = (-26.0, \ 0)$ km s$^{-1}$, which is in the range of analyses of different observation-based velocity estimates; see \citet{mobius2004,bzowski2015,schwadron2022,swaczyna2023}. The case of time-independent $\mu$ can be treated conservatively; in this case the peak of the interstellar Maxwellian VDF yields two distinct trajectories, leading to two VDF peaks at the point of interest that are associated with two distinct trajectories. One of these trajectories, which in the case of a time-independent force is characterized by having a shorter path to the target point, is called the ``direct" trajectory, the other one the ``indirect" trajectory.

\begin{figure}[ht!]
    \plottwo{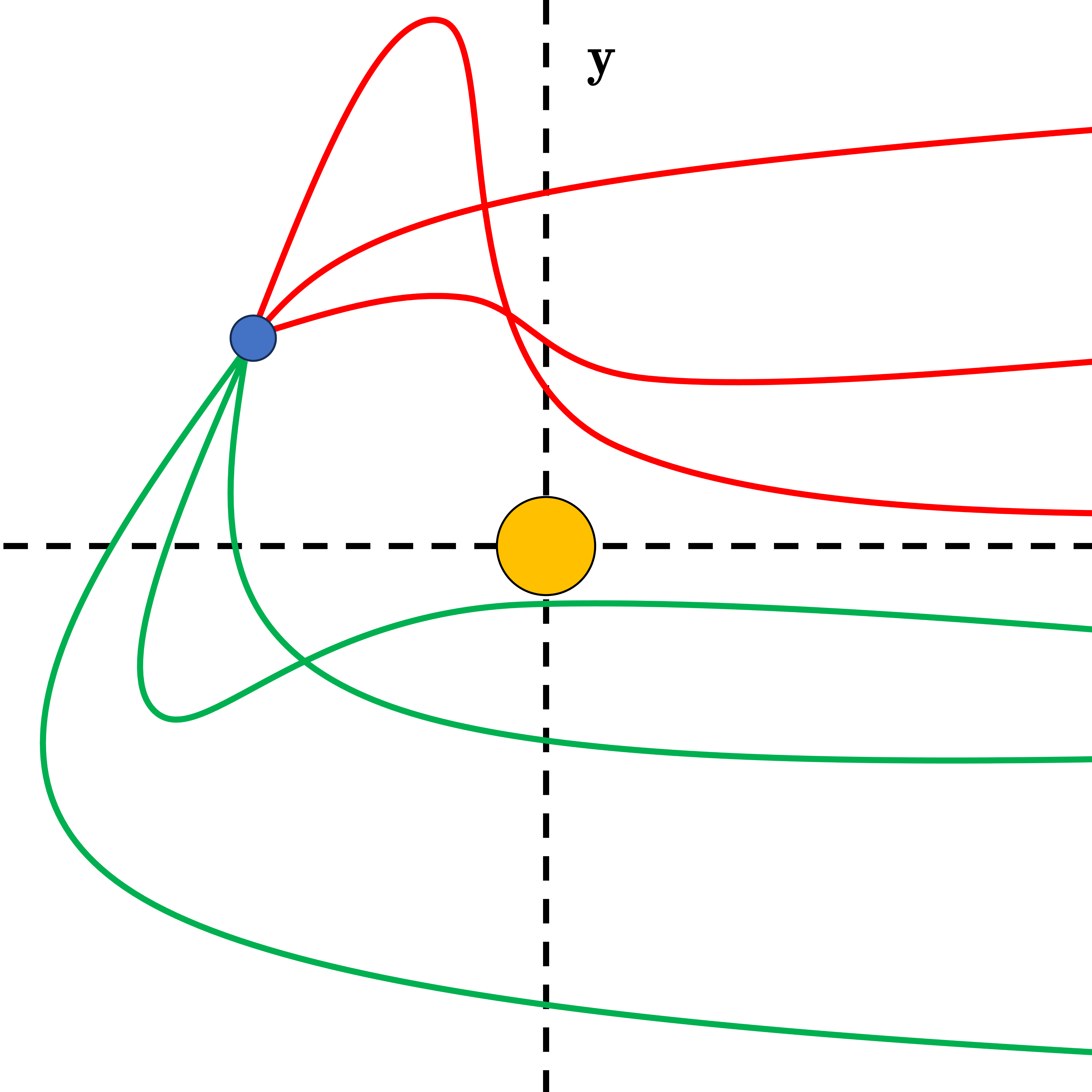}{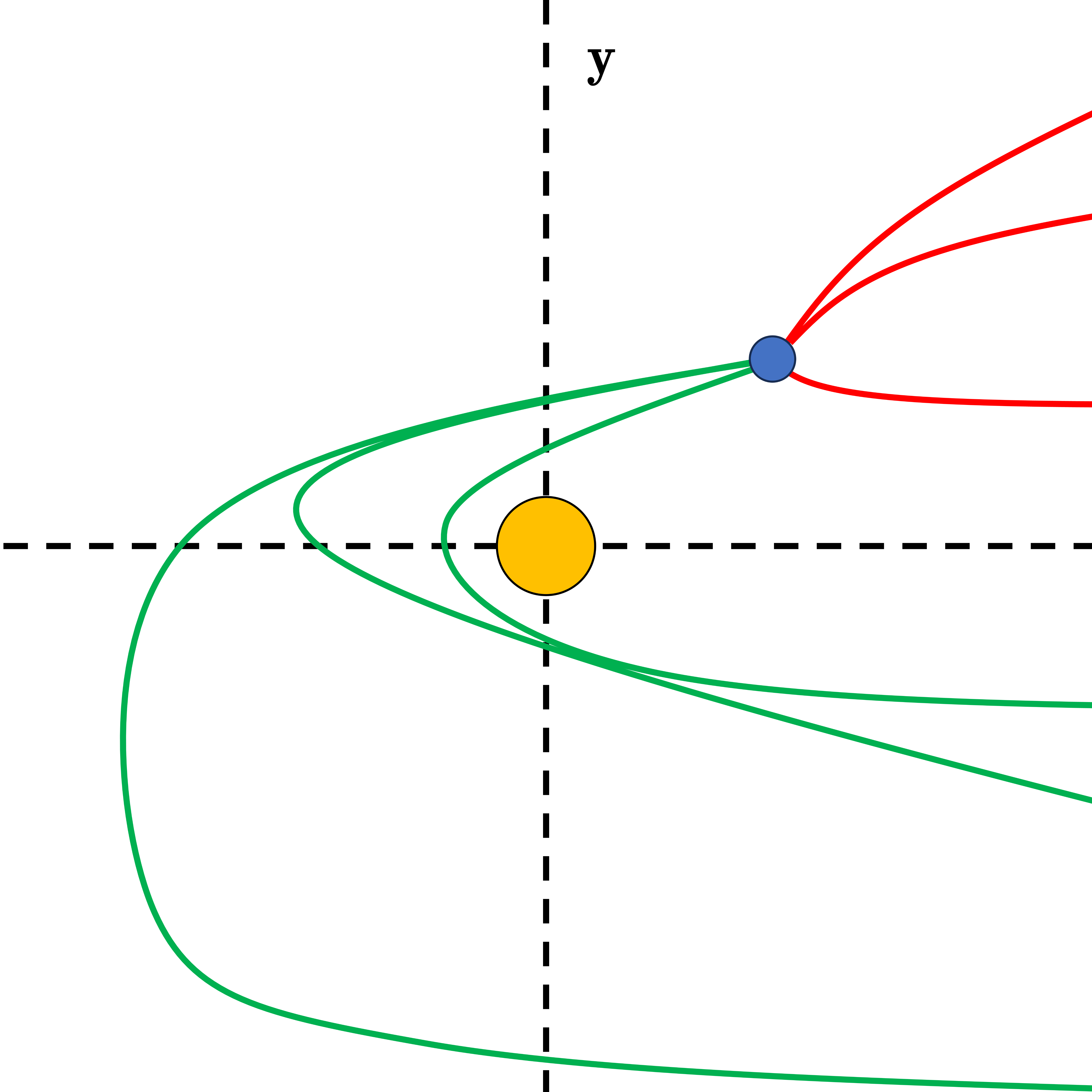}
    \caption{Sketches of representative particle trajectories, colored red for direct and green for indirect, illustrating the operational definition used in this paper for an upwind and a downwind target point (blue). Upwind is to the right; the Sun (orange) is at the origin.}
    \label{fig:trajschem}
\end{figure}

The operational definition we employ depends on the location of the spatial target point relative to the $x$-axis. If the trajectory passes through the $y$-axis on the same side of the $x$-axis as the target point before intersecting with it (for a downwind target point), or does not pass through the $y$-axis at all before intersecting the target point (for an upwind/crosswind target point), the trajectory is a ``direct" trajectory. If the trajectory passes through the $y$-axis on the opposite side of the $x$-axis as the target point before intersecting with the target point, the trajectory is defined to be an ``indirect" trajectory. Figure \ref{fig:trajschem} illustrates this definition.
\begin{figure}[ht!]
    \plotone{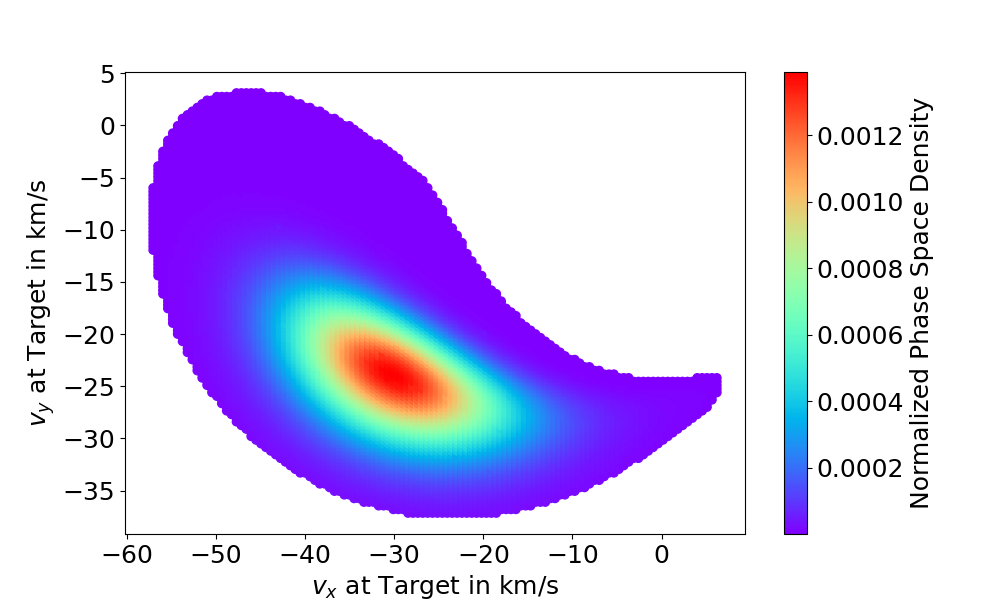}
    \plotone{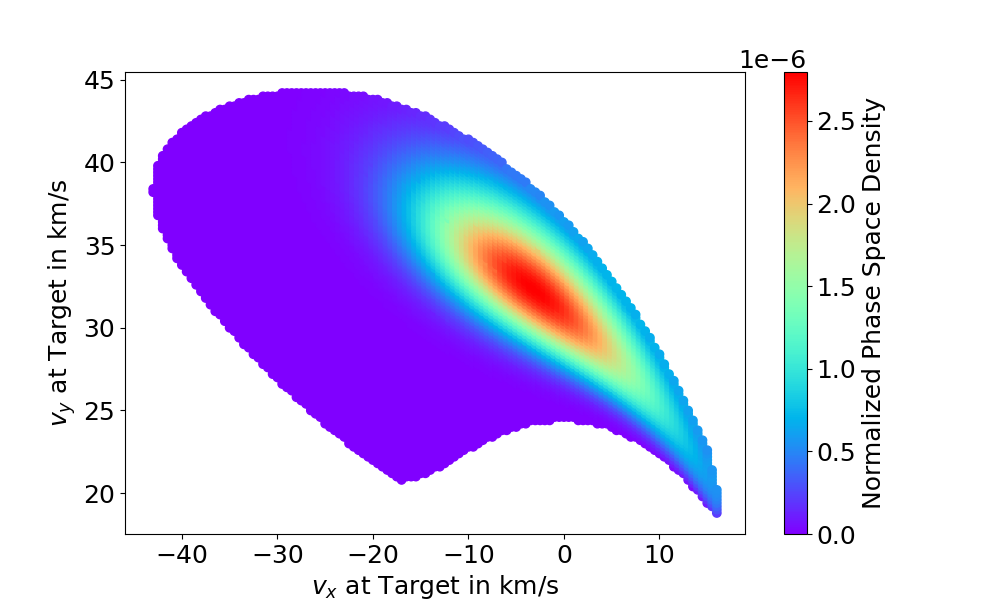}
    \caption{Direct (top) and indirect (bottom) VDF components corresponding to a sample downwind location ($-0.87,$ $0.5$) au, with a constant $\mu = 0.66$ and ionization losses (calculated according to Section \ref{sec:photoionization} below). The phase space density of each point is normalized to the LISM peak VDF value, such that the peak values of the direct/indirect cores are equal to $1$. Note that the direct and indirect components occupy notably different regions in phase space due to the different directions from which they approach this particular target point. Because of the time-independent force, these two VDF's are stationary. The edge of the VDF structure in this and following phase space plots is where the VDF value without losses reaches 0.1\% of the peak VDF value.}
    \label{fig:norpcomponents}
\end{figure}
Our study uses the warm model of interstellar neutrals, as is also employed by \citet{mueller2012} for helium. The peak of the interstellar Maxwellian (the trajectories starting with the velocity identical to $\mathbf{u}_{ISM}$) gets propagated in velocity space, defining direct and indirect peaks, and all trajectories landing in the vicinity of the direct peak are called direct trajectories as an expanded definition. All trajectories with target velocities close to the indirect peak velocity are called indirect trajectories. VDF's will be non-zero throughout velocity space, and therefore there will be contours along which the VDF is at a local minimum, delineating the boundary between direct and indirect regions. These velocity regions will henceforth be referred to as the direct and indirect components of the Maxwellian transported from infinity. 

An example of how these distributions manifest in velocity space in the time-independent case with a constant radiation pressure force representative of solar minimum is given in Figure \ref{fig:norpcomponents}.

\subsection{The Radiation Pressure Force on Neutral H}

In contrast to the time-independent case, we employ a radiation pressure force on the neutral hydrogen atoms that explicitly depends on time, neglecting the associated solar line profile function and instead assuming a flat profile. This absolves us, for the moment, from treating the dependence of the radiation pressure on the radial velocity of the neutral H atom. Even when averaged over frequency, the Ly-$\alpha$ intensity varies with the solar cycle by about $50\%$ of its maximum value \citep{lemaire2015}.

For the purpose of modeling orbits of neutral hydrogen in the region around the Sun, an idealized function $\mu (t) = 0.75 + 0.243 \cos \left(\omega\, t \right) e^{\cos (\omega\, t)}$ \citep{rucinski1995} is used for the time-dependence of radiation pressure, with $\omega = \frac{2 \pi}{T}$, and $T=3.470 \times 10^{8} \text{ s}$ the assumed duration of a solar cycle. Note that this idealized solar cycle duration is slightly less than $11$ Julian years ($3.471 \times 10^{8}$ s). This form of $\mu(t)$ has a sharp peak of $\mu = 1.41$ and dips below 1 (i.e., the net force becomes attractive) to become flat (at $\mu = 0.66$) for a period before peaking once more, which is a functional fit to the approximate appearance of the radiation pressure observationally derived during a typical solar cycle \citep{KL2018}. The peaks of the function correspond to times of solar maximum, and the minima correspond to solar minima.

\begin{figure}[ht!]
    \plotone{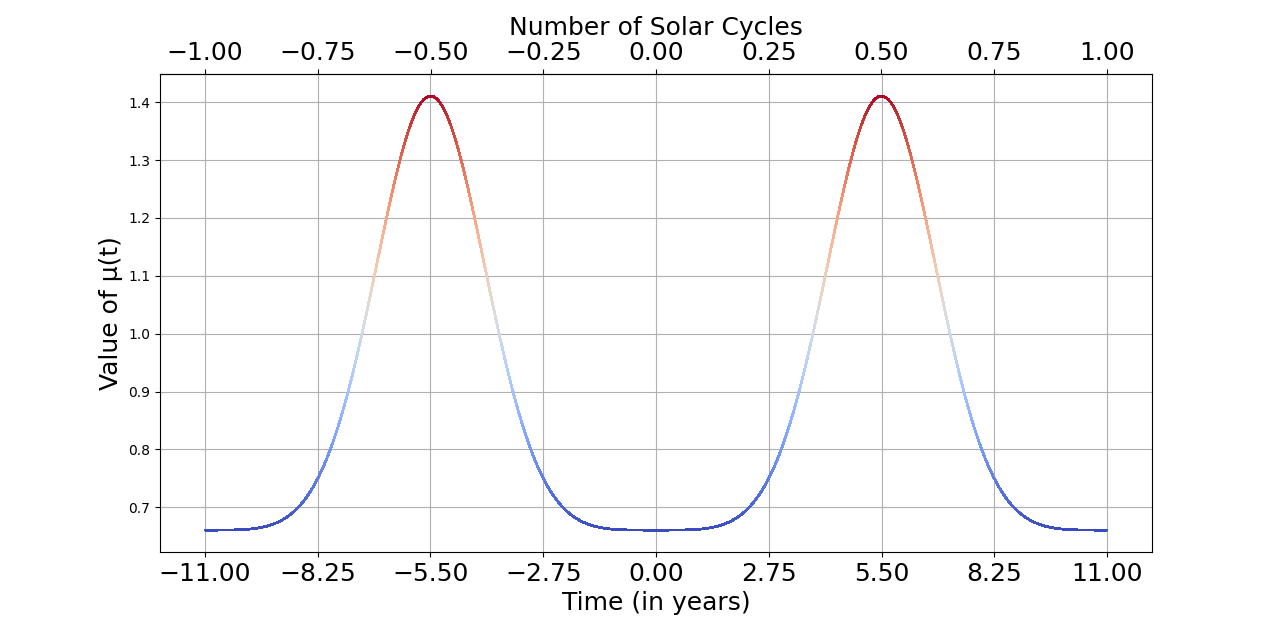}
    \caption{Plot of the function used in the code for the radiation pressure force term $\mu (t)$ as given in Equation (\ref{musimple}). The line is colored according to the value of $\mu$.}
    \label{fig:rpfunct}
\end{figure}

To assist in computational consistency, the radiation pressure function used in the code is shifted so the minimum of the function (and therefore solar minimum) occurs at $t=0$. Thus, for the actual function used in the code, $\omega t$ is replaced by $\omega t - \pi$, leading to the $\mu (t)$ function (Figure \ref{fig:rpfunct}) being written as
\begin{equation}
    \mu (t) = 0.75 - 0.243 \cos \left(\omega\,  t \right) e^{-\cos (\omega\, t)}.
    \label{musimple}
\end{equation}

All phase space plots below have their times given within the cycle before and the cycle after the minimum at $t=0$.

\begin{figure}[ht!]
    \plotone{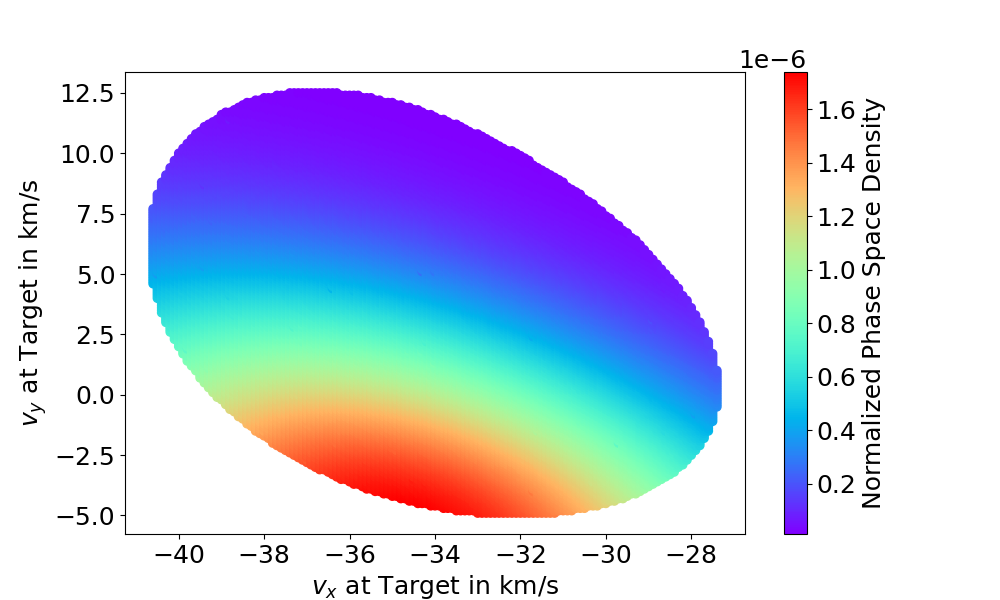}
    \caption{A VDF at a downwind spatial target point of ($-0.71,$ $0.71$) au, with $\mu$ time independent and at a value of $1.41$, the maximum value from Equation (\ref{musimple}). This target point is in the zone of exclusion, and the VDF is a remnant of what it typically is away from the exclusion zone, with many trajectories sampled from the interstellar Maxwellian unable to reach the target point. Direct and indirect trajectories are contributing to the VDF and are not clearly separated. The VDF remnant here is small, and would disappear completely already for a slightly more downwind target point.}
    \label{fig:mumax}
\end{figure}

The addition of an explicitly time-dependent radiation pressure affects the location and shape of the direct and indirect components in phase space mentioned above. Even for constant radiation pressure, the size and location of both components are changed as $\mu$ is changed. Notably, for a magnitude of the radiation pressure force where $\mu > 1$, an ``exclusion zone" exists in the downwind (and part of the upwind) direction whose locations are unreachable by trajectories thanks to the net repulsive force being experienced by the atoms, which is increasingly important in the region close to the Sun. An example of the effects of the exclusion zone on VDF for target points close to it is given in Figure \ref{fig:mumax}. This exclusion zone increases in size as the value of $\mu$ increases, or for slower atom velocities \citep{axford1972}.

\subsection{Ionization Losses} \label{sec:photoionization}

One of the considerations that must be taken into account is losses of neutral hydrogen atoms streaming in from the LISM into the heliosphere, as there are multiple mechanisms that ionize these particles. After ionization, these particles are lost from the primary neutral hydrogen population.

\begin{figure} [ht!]
    \plotone{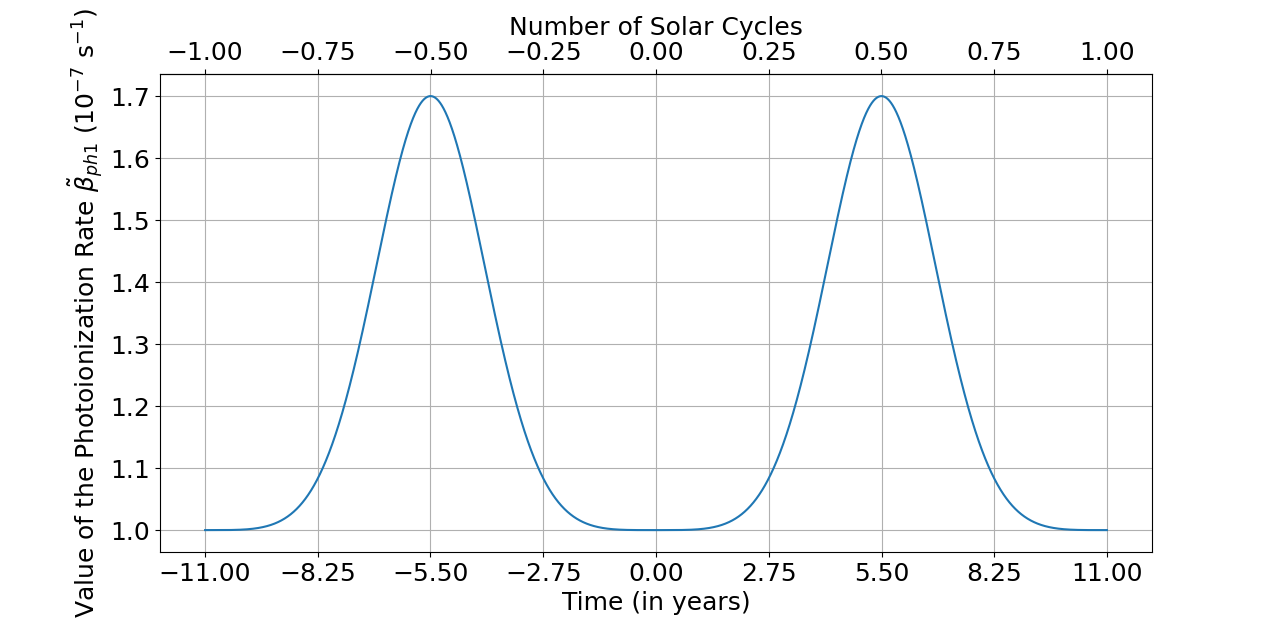}
    \caption{Plot of photoionization rate at $1$ au $\tilde{\beta}_{ph1}(t)$ as given in Equation (\ref{betasimple}).}
    \label{fig:pirate}
\end{figure}
The loss mechanisms typically considered are photoionization, electron impact, and charge exchange. In the region between Sun and termination shock, photoionization and charge exchange dominate and obey a $1/r^2$ dependence on the distance to the Sun, and hence can be treated together. Charge exchange is the primary mechanism for losses of neutral hydrogen within the termination shock \citep{sokol2019}. The ionization rate for charge exchange does not depend on time in the same way as the photoionization rate, as charge exchange depends on the solar wind which in turn depends on the solar cycle in a less orderly way \citep{sokol2019}. Thus, we have used a time-averaged value for the charge exchange ionization rate at $1$ au, approximated from the ionization rate given by \citet{sokol2019}:
\begin{equation}
    \tilde{\beta}_{cx1} = 5 \times 10^{-7} \text{s}^{-1}.
\end{equation}
When considering the $1/r^2$ dependence throughout the inner heliosphere as mentioned above, this becomes
\begin{equation}
    \tilde{\beta}_{cx} = \tilde{\beta}_{cx1} \left( \frac{r_1}{r} \right)^2
    \label{cxrate}
\end{equation}
where $r_1$ is the reference distance, in this case $1$ au. Photoionization is related to the solar EUV spectrum containing photons with energies around 13.6 eV. The rate is given by, similar to Equation \ref{cxrate},
\begin{equation}
    \tilde{\beta}_{ph}(t) = \tilde{\beta}_{ph1}(t) \left( \frac{r_1}{r} \right)^2
\end{equation}
\citep[after][]{sokol2019}
where $\tilde{\beta}_{ph1}(t)$ is the photoionization rate at $1$ au. The rate depends on the solar cycle, and for an idealized description, we make use of the function
\begin{equation}
    \tilde{\beta}_{ph1}(t) = 10^{-7} \left( 1 - \frac{0.7}{e+\frac{1}{e}} \cos \left(\omega\, t \right) e^{-\cos ( \omega\, t )} \right) \text{s}^{-1},
    \label{betasimple}
\end{equation}
\citep[after][]{rucinski1995}
which is depicted graphically in Figure \ref{fig:pirate}.
This idealized function reflects observational results about the photoionization rate of neutral hydrogen at $1$ au \citep{sokol2019}. The photoionization rate is in phase with the radiation pressure, such that there is a minimum at solar minimum and vice versa, in contrast to the assumptions used earlier by \citet{rucinski1995}.

To obtain the attenuation due to ionization losses, the local ionization rates are then integrated along the path of the neutral hydrogen atom,
\begin{equation}
    \beta_{ph} = \int_{r=\infty}^{r} \tilde{\beta}_{ph}(t)dt
    =\int_{r=\infty}^{r}\tilde{\beta}_{ph1}(t(r))  \frac{r_1^2}{v_r r^2} dr
    \label{betaintegral}
\end{equation}

and
\begin{equation}
    \beta_{cx} = \int_{r=\infty}^{r} \tilde{\beta}_{cx}dt
    =\int_{r=\infty}^{r}\tilde{\beta}_{cx}  \frac{r_1^2}{v_r r^2} dr
    \label{cxintegral}
\end{equation}
with $v_r = \frac{dr}{dt}$ being the radial velocity component (which can be positive or negative). The phase space density (PSD) at the target point is then the product of the PSD of the trajectory at our reference distance $x = 100$ au and $e^{-(\beta_{ph} + \beta_{cx})}$ \citep{bzowski2013}.

\section{Simulation Process and Relevant Code} \label{sec:code}

The Python code used to obtain results solves the equations of motion for each particle. The code employs a backtracing process to ensure all relevant trajectories ultimately end up at the same target point at the same instant in time. The specification of the time of arrival of all of the trajectories is necessary due to the time-dependent nature of the radiation pressure force. This approach was employed since, in contrast to the conservative time-independent approach \citep{mueller2012}, a time-dependent force is inherently non-conservative and therefore cannot be treated with conserved quantities. The variation of the radiation pressure force over time will lead to different sets of viable trajectories per spatial target point at each instant in time during the solar cycle.

\subsection{Backtracing Process} \label{sec:backtrace}

The 2D code starts with a 2D grid of ``initial" (in reality: final) conditions for the velocity in the $x$ and $y$ directions (labelled $v_x$ and $v_y$) equally spaced throughout each dimension in velocity space at a specific given spatial point a distance of $1$ au from the Sun at a specific time. The trajectories corresponding to each pair of final $v_x$ and $v_y$ are then traced backward in time using odeint. The trajectory is followed a sufficient time that is manually provided as an input parameter, and then the trajectory data are probed for two specific factors: whether the trajectory passes through the plane defined by $x = 100$ au, which will henceforth be referred to as the ``injection plane," and whether the velocity as it passes through this plane is contained within a (time-independent) shifted pristine LISM Maxwellian centered around the velocity $\mathbf{v}_0= (v_{x0},$ $v_{y0}) = (-26.0,$ $0)$ km s$^{-1}$ in velocity space. We consider $10^{-3}$ times the peak Maxwellian value to be the boundary of the shifted Maxwellian, which means that the velocity needs to be within a radius of $v_b \sim 26.8$ km s$^{-1}$ from $\mathbf{v}_0$. This corresponds to a thermal velocity of $v_{th} \cong 10.2$ km s$^{-1}$ (equivalent to a temperature of $T=6300$K for hydrogen in the LISM, based on \citet{heerikhuisen2014}) in the equation for the 3D Maxwellian phase space density,

\begin{equation}
    f_{ISM}(\mathbf{x},\mathbf{v}) = n_{ISM}(\mathbf{x})\, \left(\frac{1}{\sqrt{\pi} v_{th}} \right)^3 e^{-\frac{(v_x-v_{x0})^2 + v_y^2 + v_z^2}{v_{th}^2}},
\end{equation}
\citep{mueller2012}
with $v_{th}^2 = \frac{2k_B T}{m_H}$. A normalization, namely dividing out the term in front of the exponential to leave

\begin{equation}
    f(\mathbf{v}) = e^{-\frac{(v_x-v_{x0})^2 + v_y^2 + v_z^2}{v_{th}^2}},
    \label{normpsd}
\end{equation}
is done such that the value of the function decreases from $1$ at $\mathbf{v}_0$ to $10^{-3}$ at $|\mathbf{v}-\mathbf{v}_0| = v_b$; the $10^{-3}$ contour being a sphere, and in 2D, a circle. This choice of a contour is such that the set of trajectories we consider accounts for $99.95\%$ of the total distribution of primaries by density based on our assumption of a Maxwellian.

In addition, the code first checks after calculating the trajectory if any point of the trajectory passes through the Sun or its surface, in which case the trajectory is no longer considered.

If these conditions are met, the final values for the velocity components of the trajectory are saved, as well as the attenuated value of the normalized phase space density. This allows generation of phase space plots and study of their properties.

The time resolution is split up into two regimes to improve the efficiency of the code --- a close (to the Sun) regime and a far regime. The close regime uses a finer resolution in time to help minimize issues that arise in the integration of the ionization rates as given in Equations (\ref{betaintegral}) and (\ref{cxintegral}), as turning points in the value of the radial distance ($v_r = 0$) can produce issues with the integration process thanks to the value of $dr$ increasing and decreasing simultaneously over a single interval if the resolution is not fine enough. However, as it is not anticipated this will be an issue once the trajectory is outside the region close to the Sun, the time interval can be increased with little concern, as the velocity will also largely stay the same from one interval to the next, especially as the trajectory approaches the injection plane. A coarse time resolution in the close regime is also liable to miss the event that the trajectory's path intersects the Sun and hence should be discarded. This effect is hard to predict, and adds some weight to the choice of time resolution in this regime. The choice of both resolutions is made manually to ensure both proper determination of losses near the Sun as well as proper calculation of ionization rates.

\subsection{Details on the odeint Method} \label{sec:odeint}

The main method used for calculation of trajectories within the code is the odeint method from the scipy.integrate library \citep{scipy2020}. This method allows for solving for trajectories that is consistently accurate regardless of the time resolution, allowing the solver to be run at high resolutions in the region close to the Sun (within a radius of about $5$ au) and at lower resolution farther from the Sun for reasons mentioned above.

The odeint method solves the system using the lsoda solver from FORTRAN's odepack library, which solves initial value differential equation problems while adjusting between a stiff (Backwards Differentiation Formula) and nonstiff (Adams) method automatically depending on the nature of the problem at a given location in space and time. Since the solver handles the explicit problem of the form $\frac{dy}{dt} = f(t,y)$ for $t$ being the independent variable and $y$ the dependent variable \citep{hindmarsh1992}, the second order equation for the force is split into a system of first order equations,
\begin{equation}
    \dot{x}_i = v_{x_i} 
\end{equation}
\begin{equation}
    \dot{v}_{x_i} = -G M_{\odot}\frac{x_i}{r^3} (1 - \mu(t)) = a_{x_i}
\end{equation}
for each spatial dimension being used with $x_1=x$, $x_2=y$, and $x_3=z$ in Cartesian coordinates. In three dimensions, this gives a system of $6$ differential equations, coupled into sets of $2$ for each dimension. The time dependence of $\mu (t)$ is allocated to a separate function input into the odeint function, such that the form can easily be changed from test to test if needed, without editing the base differential equation set. In this case, the code is employed in 2D, which leaves $4$ equations in $2$ sets. Thus, it is possible to accurately determine the trajectories of particles in this time-dependent force environment by backtracing using an externally generated array of decreasing values of the time variable $t$ that odeint takes as input.

\section{Discussion of Results} \label{sec:results}

\subsection{Differences in Direct and Indirect Components for Time-Dependent Case}

\begin{figure}[ht!]
    \plotone{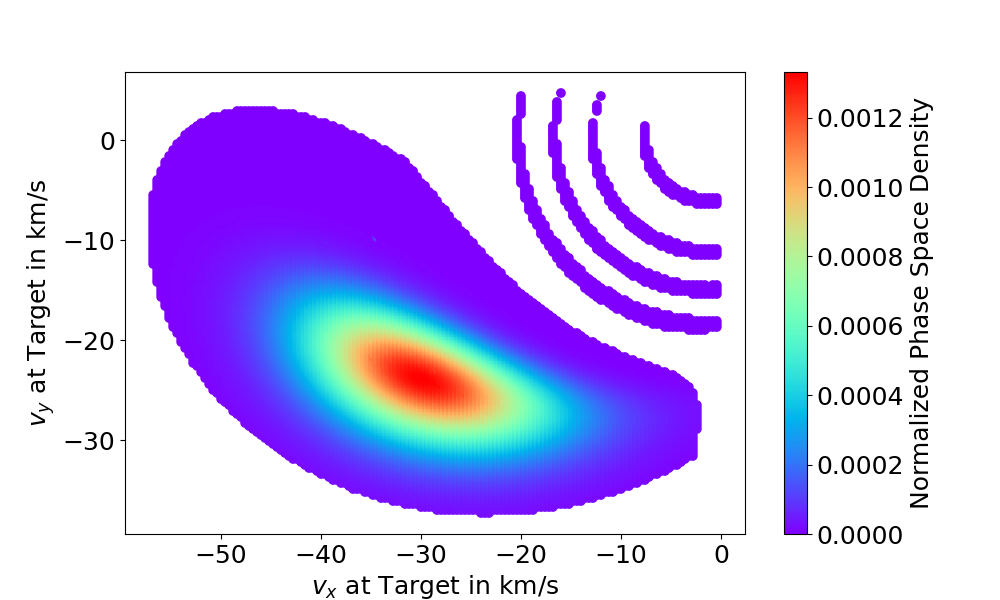}
    \plotone{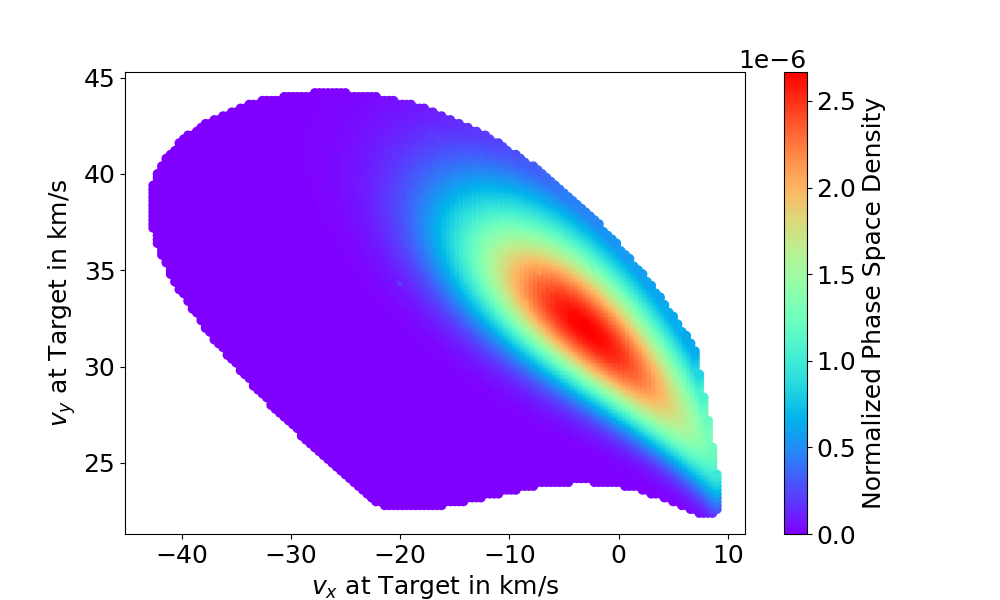}
    \caption{Phase space plots corresponding to the direct (above) and indirect (below) components at the same spatial target point ($-0.87,$ $0.5$) au as in Figure \ref{fig:norpcomponents}, at a time of $t = 0.13$ years ($\mu$ $= 0.66$ still). The time-dependent radiation pressure force term as given in Equation (\ref{musimple}) is used to arrive at these VDF. The shape, size, and location are all different from the corresponding components shown in Figure \ref{fig:norpcomponents} with the same color bars.}
    \label{fig:componentsrp}
\end{figure}

The direct and indirect components of the hydrogen velocity distribution function are notably different from the case of a time-independent scenario. For example, the plots in Figure \ref{fig:componentsrp} give the two components at the same spatial and temporal point as the components in Figure \ref{fig:norpcomponents}, with the properties differing due to the addition of time-dependence to $\mu (t)$. The components themselves are time-dependent, and in the case of generally downwind locations, whether the components are present or not depends on the phase in the solar cycle in which one observes that location. If the location is in the exclusion zone corresponding to values of $\mu$ close to the maximum, the components will disappear entirely during the period of exclusion, reappearing when the force becomes attractive once more, with the direct component appearing first. Once sufficient time has passed since the force has become attractive, the indirect component will appear in phase space as well for these points. The delay is caused by the finite timescale of any trajectory's path into the region in the vicinity of the Sun, as the force must be attractive for a period of time for an indirect trajectory to arc around the Sun on the downwind side after passing the $x=0$ plane.

Due to the nature of the force increasing and then decreasing in its magnitude based on the form of Equation (\ref{rpforce}) even when in the attracting phase, the direct and indirect components evolve and move to regions that correspond to higher particle velocities as the force stays attractive before moving back toward slower velocities. In the case of a generally downwind location within the exclusion zone as mentioned before, both components will eventually shrink and disappear, but in the case of other locations, only the indirect component as defined previously will disappear at some point after the transition from an attractive to a repulsive force. The reason the indirect component disappears is because as the force grows less attractive and switches to being repulsive, the range of impact parameters that could result in an indirect orbit slowly decreases and thins, approaching $0$, at which point there are no longer any impact parameters that lead to viable indirect orbits during the repulsive regime.

While mostly affecting downwind locations, the exclusion zone itself does generally extend into the upwind direction for any repulsive force from the Sun. However, due to the parameters of the simulation, including the form of the net force and the radial distance of spatial points studied here, all upwind points that are considered are outside of the corresponding upwind exclusion zone for the entire cycle of $\mu (t)$. The transition to a repulsive force does not immediately get rid of an indirect component, as any trajectory that has already passed through the exclusion zone to some degree can still be repulsed toward a target point if it has the appropriate velocity at the time of the transition. Thus, the indirect component of this variety lingers for a period after the transition from an attractive to a repulsive force before disappearing.

\begin{figure}[ht!]
    \plotone{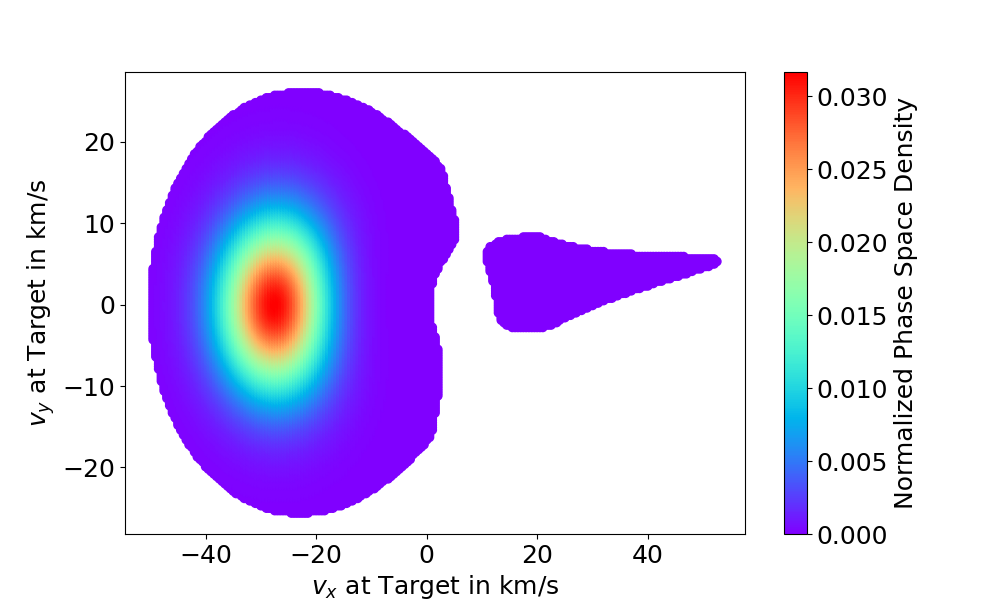}
    \plotone{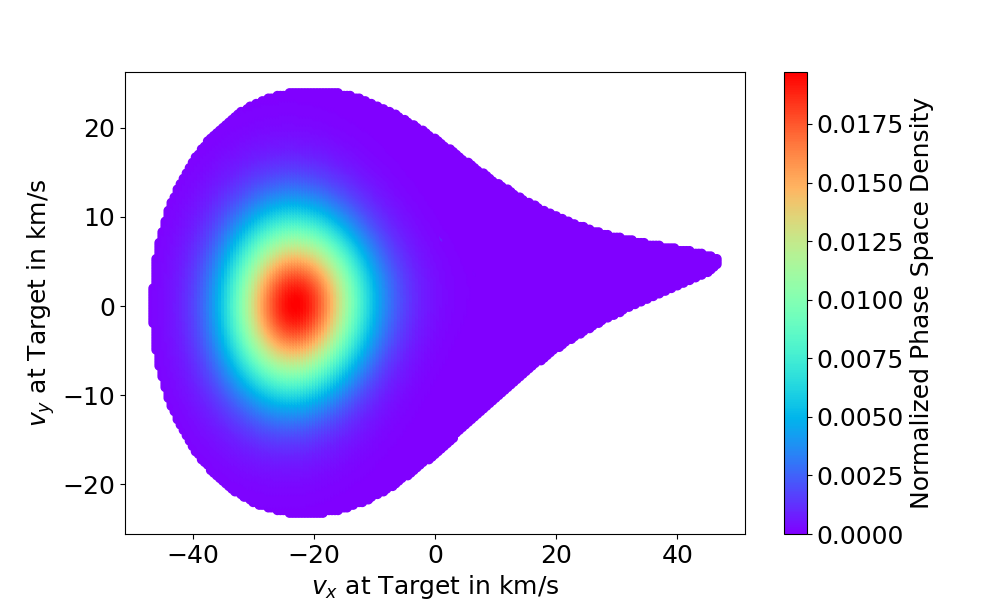}
    \caption{Population in phase space corresponding to a spatial target point at ($0.995$, $0.098$) au, first at a time before maximum repulsion while in the repulsive regime, with $\mu = 1.26$ at $t = 4.63$ years (top), then at the time of maximum repulsion when $\mu = 1.41$ at $t = 5.5$ years (bottom). The repulsive indirect component is present in both plots, in the top plot as a completely independent structure and below as an offshoot of the direct component.}
    \label{fig:repulsiveindirect}
\end{figure}

\begin{figure}[ht!]
    \plotone{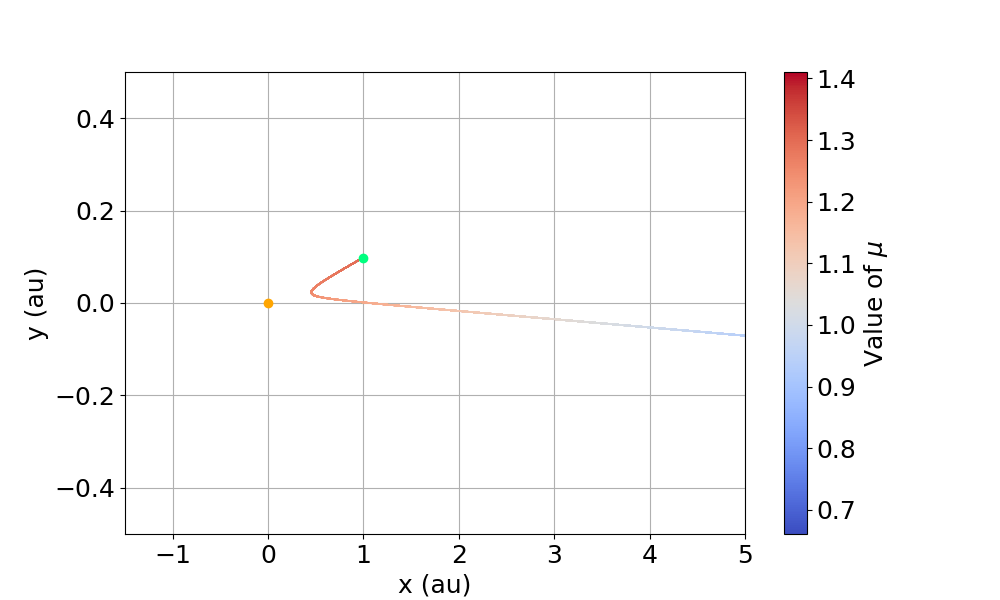}
    \caption{Individual orbit drawn from the repulsive indirect component at $t = 4.76$ years, at the same spatial target point as Figure \ref{fig:repulsiveindirect} (green dot). The area of the plot is zoomed in on the region near the Sun (orange dot), and the color bar for this and all individual orbit plots matches the coloring in Figure \ref{fig:rpfunct}. The trajectory has a final velocity of $\mathbf{v}_0 = (24.2, 2.85)$ km s$^{-1}$, has a normalized phase space density value of $f \approx 2.59 \times 10^{-11}$, reaches its perihelion at a distance of $0.45$ au from the Sun, and takes about $20$ years to reach the target point from the injection plane, which will be referred to in further plots as the ``age" of the trajectory.}
    \label{fig:repulsiveindorb}
\end{figure}

In the case of target points that are sufficiently upwind, during repulsive phases there can appear an indirect component that is an equivalent to the indirects in a time-independent repulsive case \citep[e.g.][]{axford1972}. This type of trajectory will be referred to as a ``repulsive indirect" trajectory. The repulsive indirect component corresponds to trajectories that reach a perihelion before being reflected back to the target point by the overall repulsive force present during the time of perihelion passage. It is either connected to the direct component or separate from it depending on the time at which it is observed. As the force grows less repulsive and switches to attractive, this variant of the indirect component disappears in favor of the other typical flavor of indirect component. An example of this kind of indirect component is given in Figure \ref{fig:repulsiveindirect}, with an example of an individual trajectory corresponding to those found in this type of indirect component given in Figure \ref{fig:repulsiveindorb}. This trajectory was obtained by adapting the code to trace back one phase space point in time. The initial conditions for these individual orbits are drawn from phase space plots. The phase space data of the trajectory is retained at each time point, and the trajectory is plotted in visual space with its path colored by the value of $\mu$ at each time point.

The relative timescale of the travel time of a particle on the trajectory and the period of the force oscillation create conditions such that the indirect component corresponding to an attractive force disappears. If the timescale for the oscillation was faster, or if the attractive period was longer relative to the repulsive period, this indirect component would be allowed to survive since trajectories could traverse the exclusion zone before they become excluded. However, there is a delicate balancing in this case that results in the behavior demonstrated here.

The exact VDF behaviors, specifically with regard to the location of the center of the core of each VDF component, its shape, and the overall ``movement" (trace in velocity space) of the VDF structure vary between spatial locations. However, the characteristic behaviors of the direct component appearing or being present at all times for all upwind/crosswind and a range of downwind locations are consistent across all spatial locations considered, as are the behaviors of the indirect component appearing during the attractive regime. The two components also consistently move in phase space and change shape throughout the attractive period.

\begin{figure}[ht!]
    \plotone{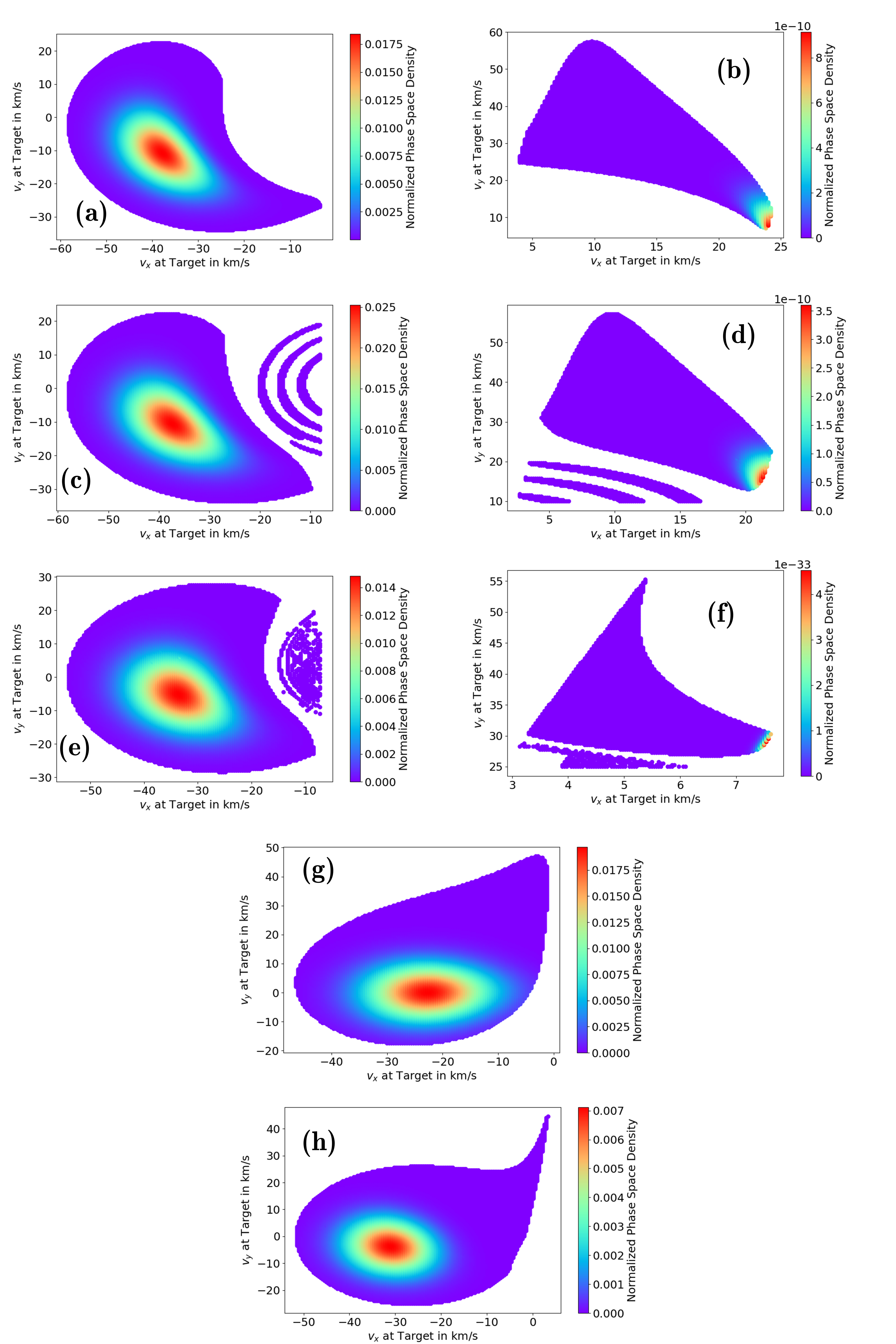}
    \caption{Time sequence of VDF at the same near-crosswind location ($0.087$, $0.996$) au. Panels (b), (d), and (f) are the indirect components; all others are direct. For reference, (a) and (b) are the time-independent VDF at $\mu=0.66$ (the minimum radiation pressure according to Equation (\ref{musimple})). The remainder of the figure represents four time instants throughout the cycle of the oscillation of $\mu (t)$ as given in Equation (\ref{musimple}). (c) and (d) are the direct and indirect components when $\mu$ is minimized at $\mu = 0.66$ at $t = 0$ years (force is maximally attractive), (e) and (f) are the components with $\mu=1$ (at $t = 3.82$ years) and increasing (changing out of the attractive regime), (g) is the direct component with $\mu$ maximized at $\mu = 1.41$ at $t = 5.5$ years (maximally repulsive), and (h) is the direct component with $\mu=1$ (at $t = 7.17$ years) and decreasing (changing back into the attractive regime after a period of repulsion). Note there is no indirect component past a certain point when $\mu > 1$, namely in (g) and (h).}
    \label{fig:timeseq}
\end{figure}

These behaviors are illustrated in the time sequence given in Figure \ref{fig:timeseq}, which portrays the available components at various times in the oscillatory cycle of $\mu$ for a close to crosswind location. Here it can be seen that the components themselves are constantly changing, and that at the force free point while switching to repulsive, the indirect component is becoming smaller as a result of the force becoming less and less attractive. The direct component continues to change shape, even starting to produce a tail as $\mu$ decreases back below $1$. Also worth noting is that the indirect component in the attractive period originally forms by detaching from the direct component as the force switches from repulsive to attractive.

\begin{figure}[ht!]
    \plottwo{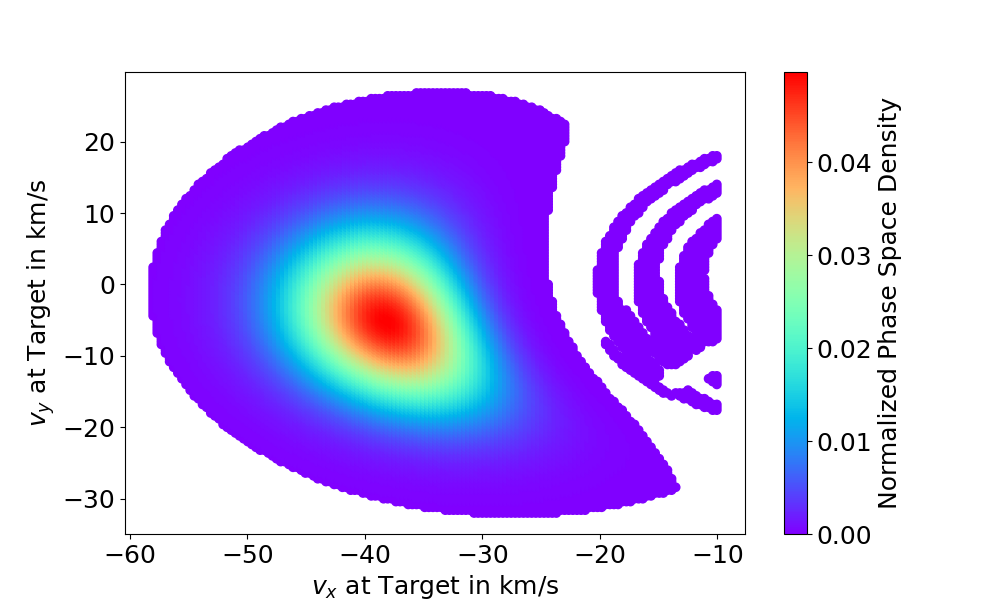}{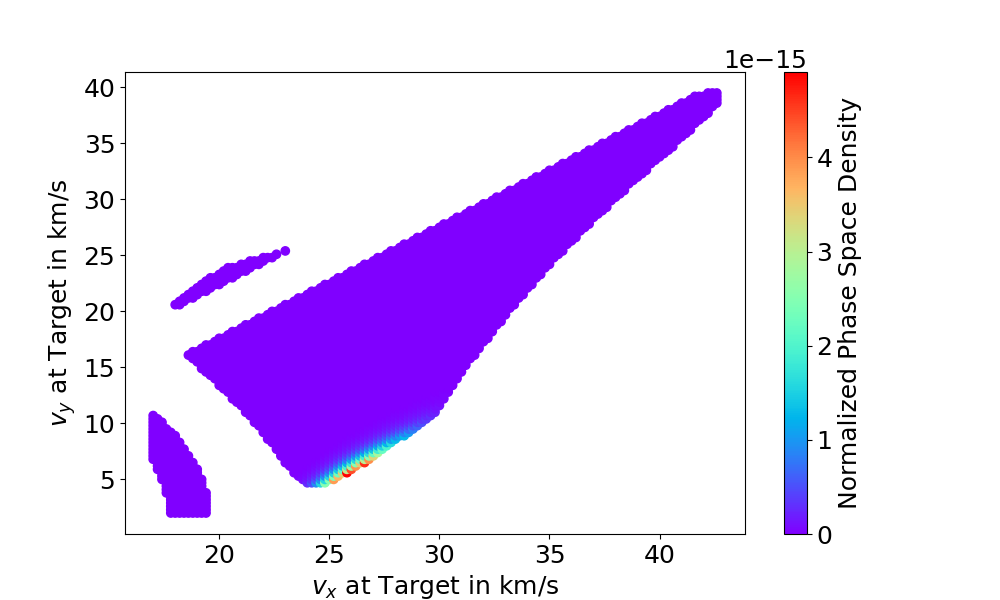}
    \caption{Direct (left) and indirect (right) VDF components taken at an upwind location at $\mathbf{r} = (0.71  \text{, } 0.71)$ au at $t = 0.13$ years. These are taken at the same temporal point as the plots in Figure \ref{fig:componentsrp}, but differ in location and size due to the different spatial target point.}
    \label{fig:dirindirupwind}
\end{figure}

\begin{figure}[ht!]
    \plotone{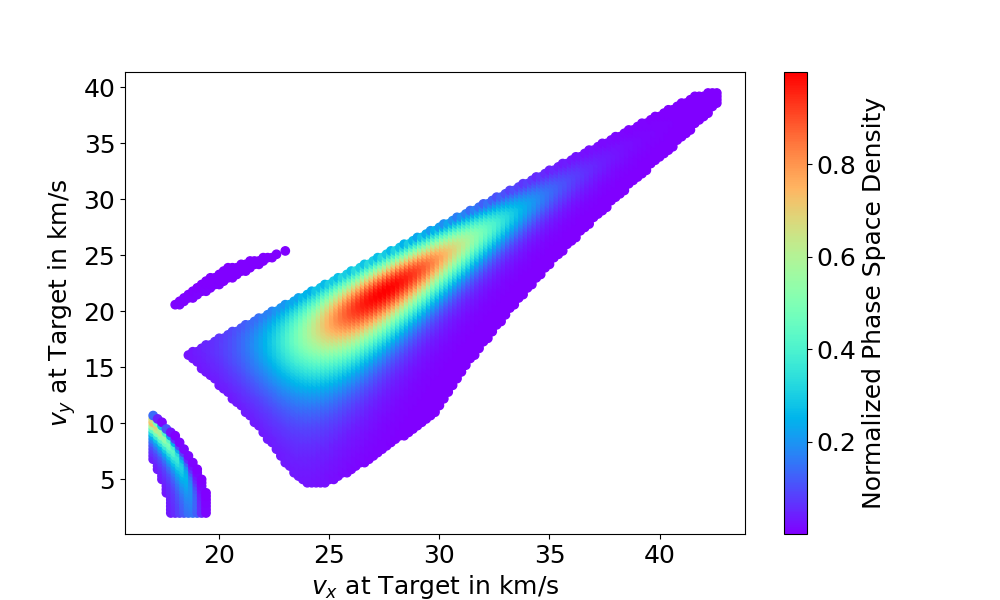}
    \caption{Plot of the indirect component from the same upwind location and temporal point used in Figure \ref{fig:dirindirupwind} (right), but without accounting for ionization losses. Note that the core is located closer to the center of the structure, in contrast to the case where losses are included.}
    \label{fig:upwindnopi}
\end{figure}

To demonstrate the effects of ionization, we also present VDF with ionization losses switched off. Ionization generally affects the direct components to a lesser degree, such that the location of the core and the fall-off in value outside of the core is similar in structure to that of the case without ionization. The maximum value of the core of the direct component is generally $< 10 \%$ depending on the instant in time the VDF is taken, with $100\%$ being the peak value when no ionization losses are considered. The magnitude of the attenuation can be greater, though, in the repulsive regime when trajectories linger in the region around the Sun for longer. The indirect component, across all spatial target points and times in the solar cycle, experiences even more significant influence from ionization across the board due to how long indirect orbits stay in the region surrounding the Sun. The preferential nature of ionization can also lead to shifting of the core of the Maxwellian within the structure due to differences in attenuation, as can be seen in Figure \ref{fig:dirindirupwind}, where the indirect component contrasts the case in Figure \ref{fig:upwindnopi}, which does not consider ionization losses.

\subsection{Manifestation of Pseudo-Bound Orbits in Phase Space} \label{sec:pb}

\begin{figure}[ht!]
    \plottwo{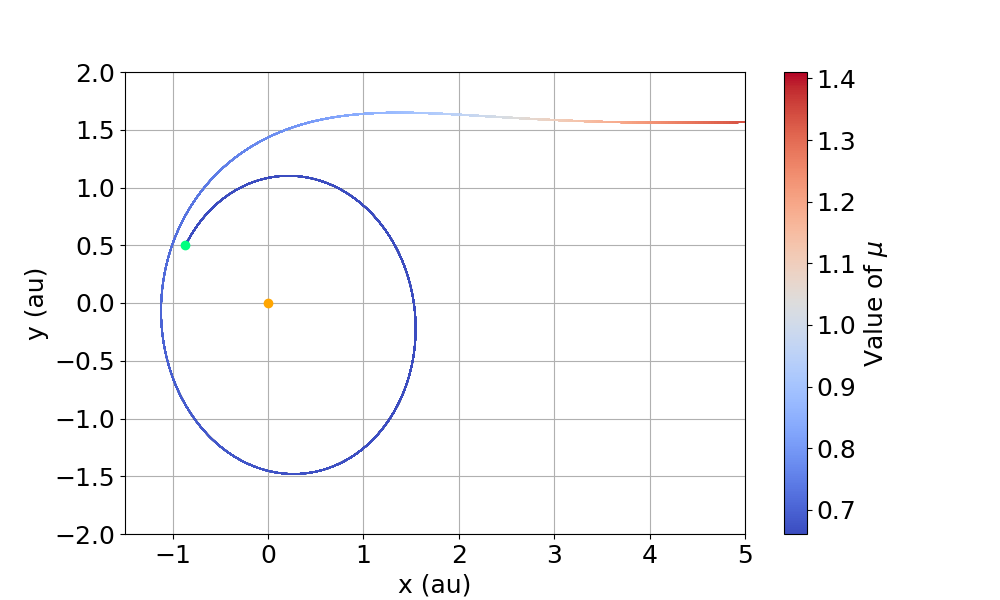}{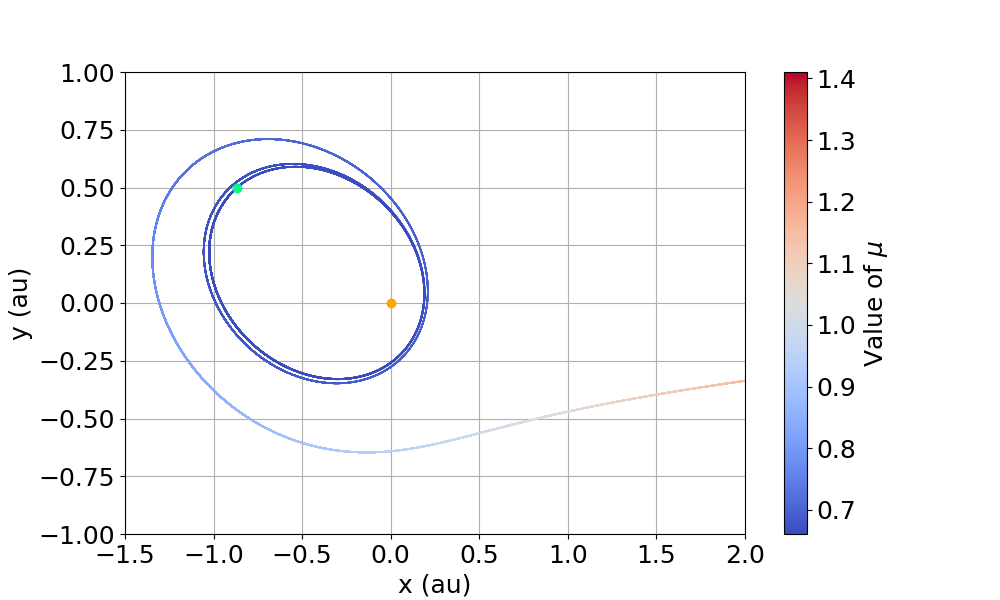}
    \caption{Individual orbits generated from the code, with a spatial target point located at $\mathbf{r} = (-0.87$, $ 0.5)$ au (green dot) at time $t = 0.13$ years, corresponding to $\mu = 0.66$.  The left trajectory has a final velocity of $\mathbf{v}_0 = (-10.0,$ $-16.5)$ km s$^{-1}$. This trajectory has a normalized phase space density value of $f = 2.32 \times 10^{-19}$, reaches its perihelion at a distance of $0.999$ au from the Sun, and has an age of about $36$ years. The right trajectory, which notably passes through the $x=0$ plane multiple times before reaching the target point, has a final velocity of $\mathbf{v}_0 = (9.0,$ $6.0)$ km s$^{-1}$ and $f = 2.02 \times 10^{-92}$, perihelion $0.18$ au, with age $26$ years.}
    \label{fig:indorb1}
\end{figure}

In the case of a central force from the Sun that is time-independent, we can consider three conserved quantities throughout the orbit of a neutral hydrogen particle: the angular momentum, the energy of the particle, and the Laplace-Runge-Lenz vector \citep{mueller2012}. In the case of a time-dependent central force, the energy and the Laplace-Runge-Lenz vector are no longer conserved. A good way of visualizing this non-conservation of energy is imagining a particle moving purely in the negative $x$-direction from far that approaches the Sun with a low impact parameter. If the particle initially experiences a repulsive force, the particle will slow as it approaches the Sun - however, if this force switches to being attractive while the particle's potential energy is nearly maximal, specifically when the particle is close to the perihelion of its orbit during a period of repulsion, the particle can become trapped in a bound orbit, entering a state with a much lower energy than before. This is due to the potential energy quickly becoming negative before the kinetic energy has time to adjust accordingly. This switching of the force can be seen in Figure \ref{fig:indorb1}, where the transition from red to blue indicates the force switching from repulsive to attractive.

This interesting case begets a particular circumstance where a time-dependent central force that alternates between repulsive and attractive can lead to the formation of ``pseudo-bound" orbits - that is to say, a particle loses energy in a process similar to that described above and ends up orbiting the Sun a number of times, during which it could potentially reach a target point. In principle, there are any number of these pseudo-bound orbits existing in the region around the Sun at any given time in the attractive period, but the observable population through this code is limited to primary neutral hydrogen thanks to the enforced conditions of the recorded trajectories belonging to the Maxwellian centered around $v_{x0} = -26$ km s$^{-1}$ at the injection plane and the finite backtracing time given as input.

\begin{figure}[ht!]
    \plotone{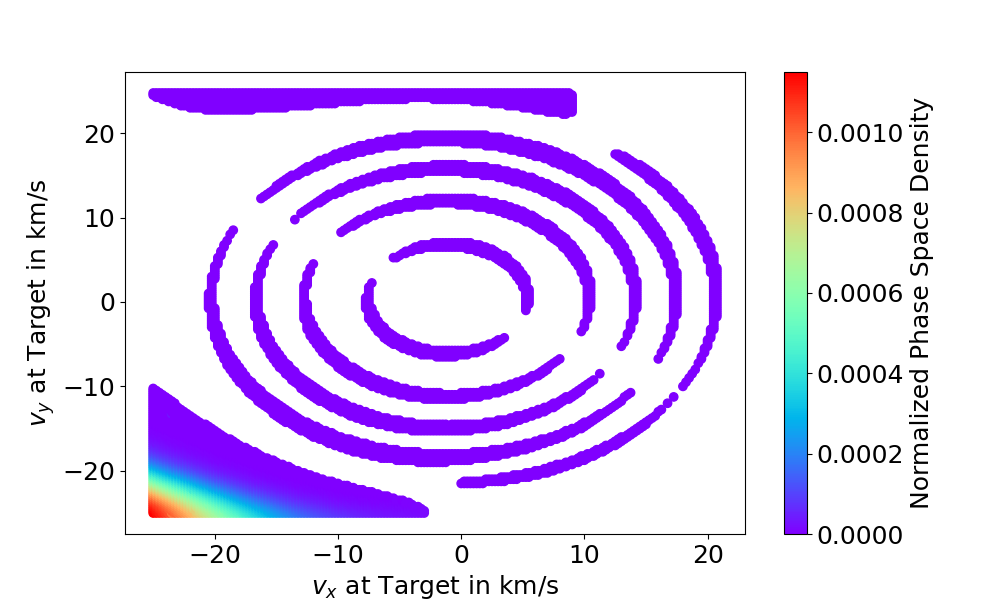}
    \plotone{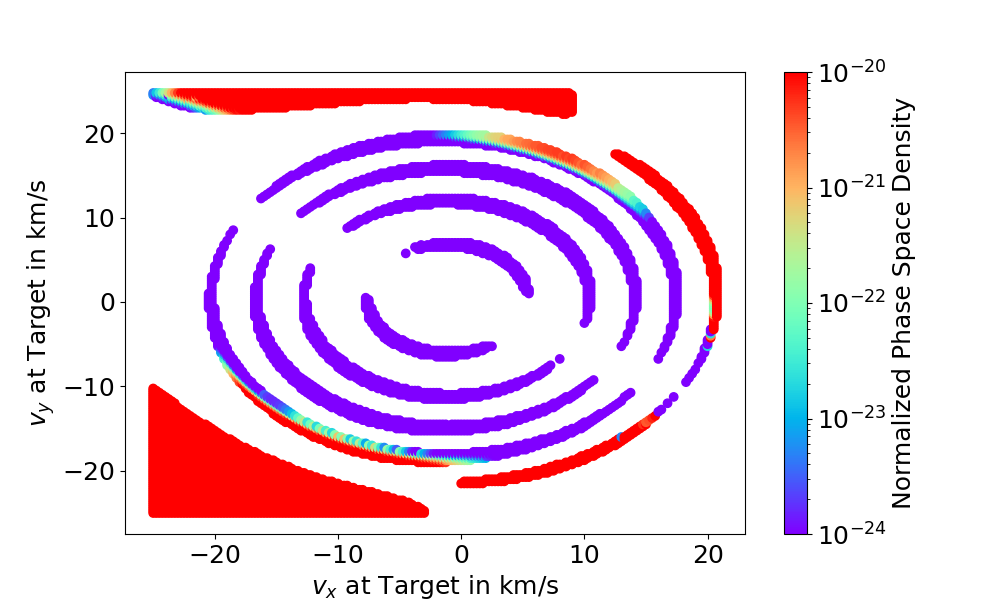}
    \caption{VDF plots at target point ($-0.87$, $0.50$) au at time $t = 0.13$ years ($\mu = 0.66$), just after a time of maximal attraction at $t = 0$. Ionization is included. Both individual orbits in Figure \ref{fig:indorb1} draw from points on this plot. The plot focuses on the fine structure in the low velocity regions. The VDF is plotted both on a linear color scale, and a logarithmic one for which values below $10^{-24}$ are all plotted as purple dots. The ring structures are present, and it is clear from the logarithmic plot that structures farther out from the center (fewer round trips) are attenuated less. Note that all log plots given in this paper span four orders of magnitude, for the sake of consistency in range.}
    \label{fig:center}
\end{figure}

\begin{figure}[ht!]
    \plottwo{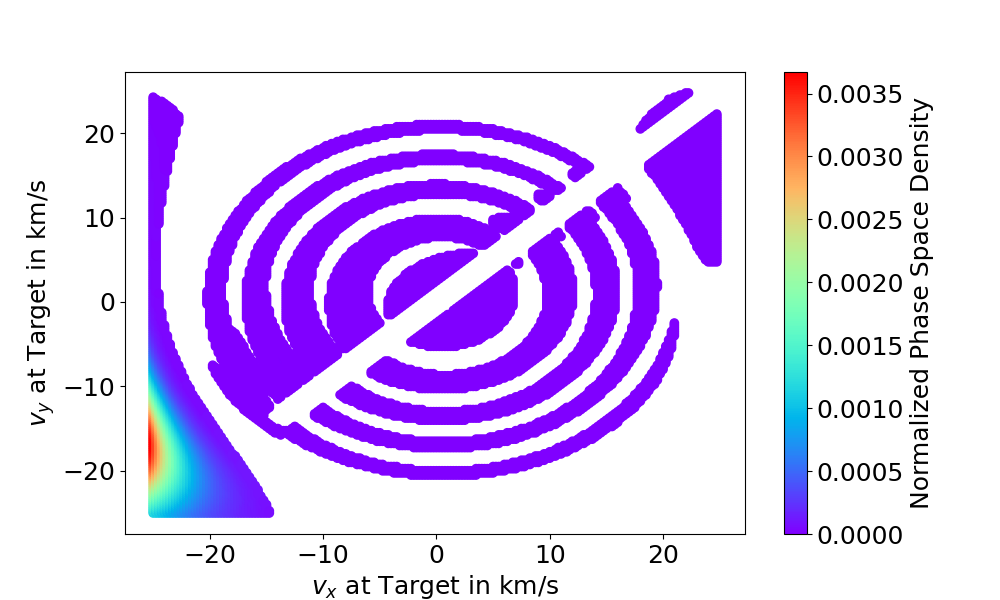}{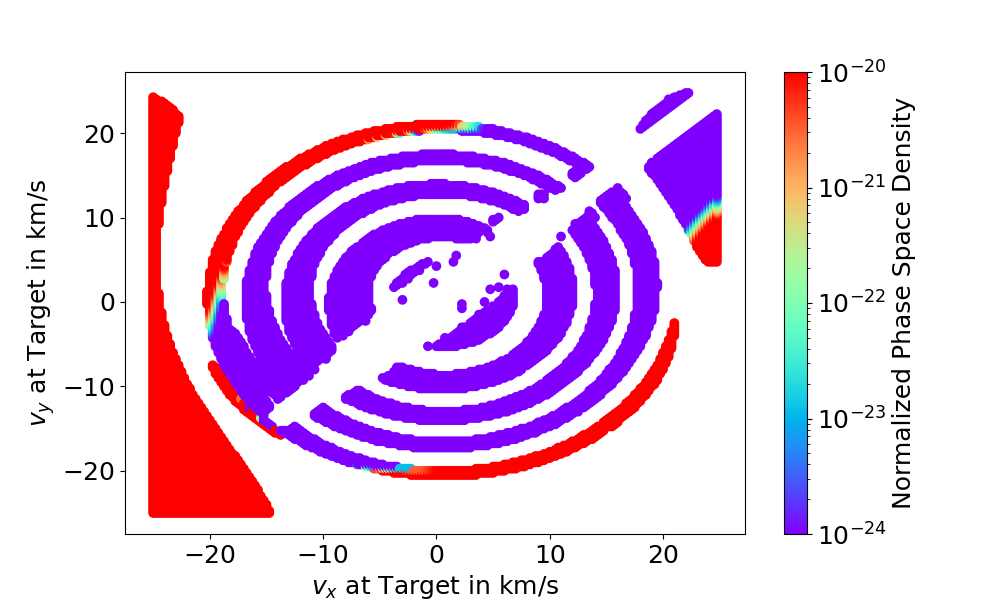}
    \caption{Center region of the VDF at the same spatial and temporal point as in Figure \ref{fig:dirindirupwind}, which is the same temporal point as the plot in Figure \ref{fig:center}. Note that the outer structure arc is incomplete here and in Figure \ref{fig:center}.}
    \label{fig:centerupwind}
\end{figure}

\begin{figure}[ht!]
    \plottwo{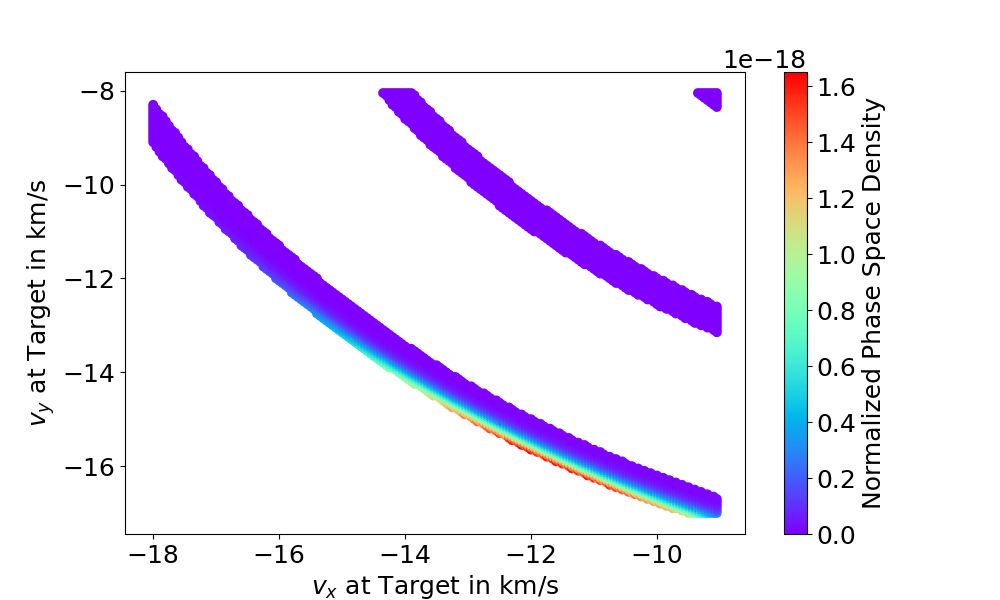}{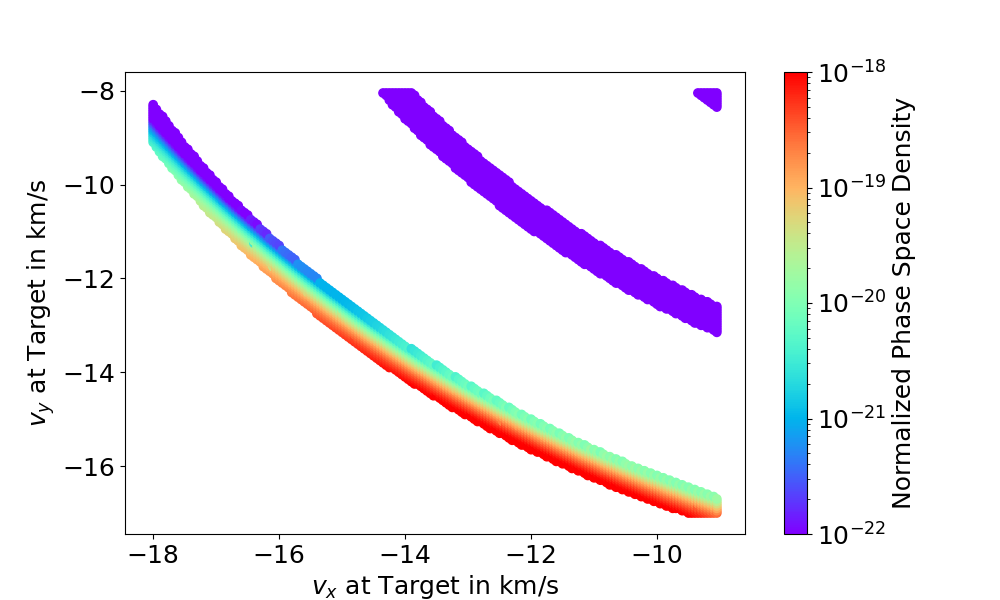}
    \caption{Zoomed-in view of a portion of the ring structures in Figure \ref{fig:center}, demonstrating the width of the structures and the variation within them. Linear color scale on the left; logarithmic one on the right.}
    \label{fig:centerzoom}
\end{figure}

\begin{figure}[ht!]
    \plottwo{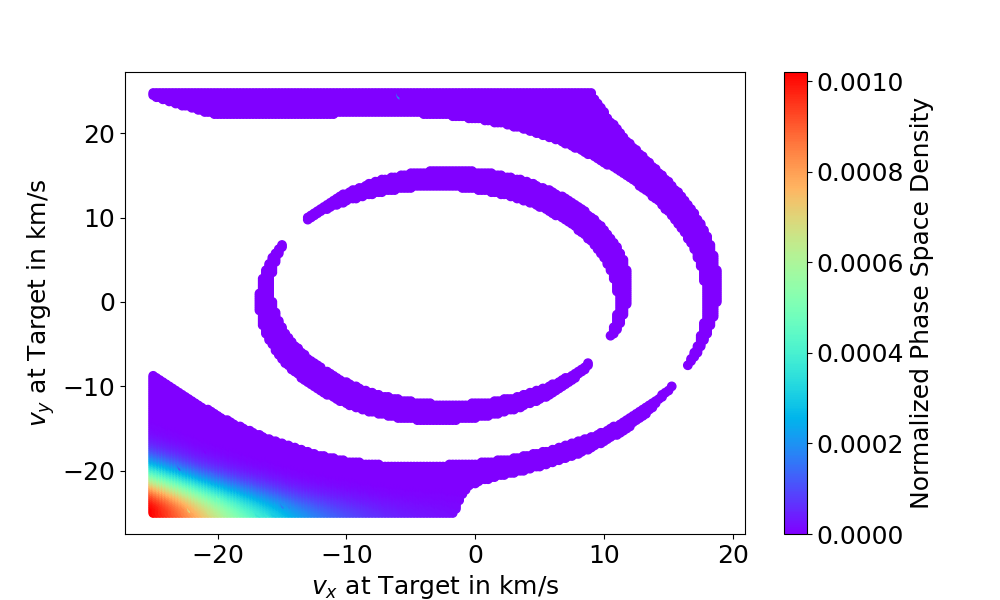}{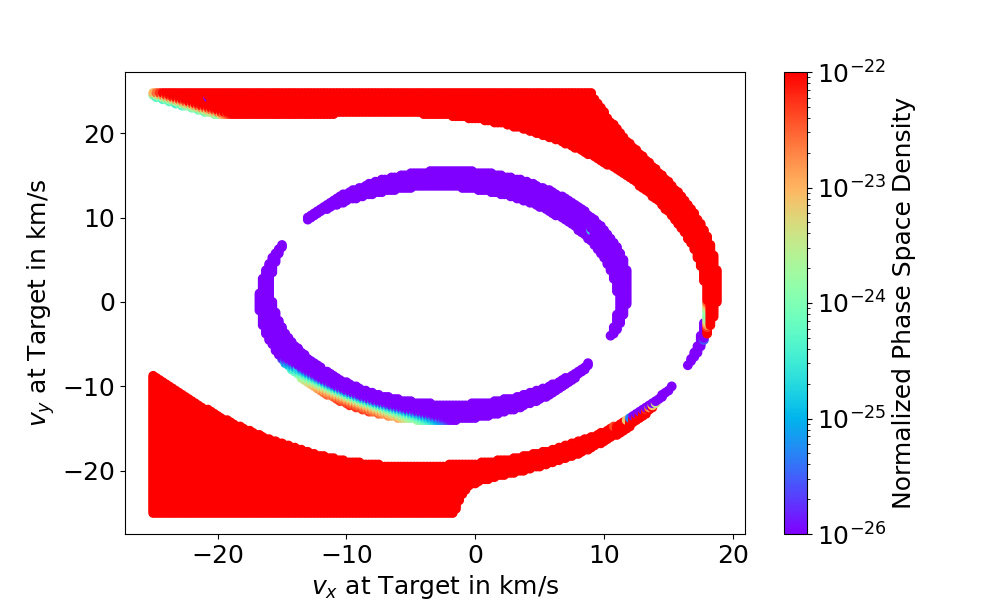}
    \plottwo{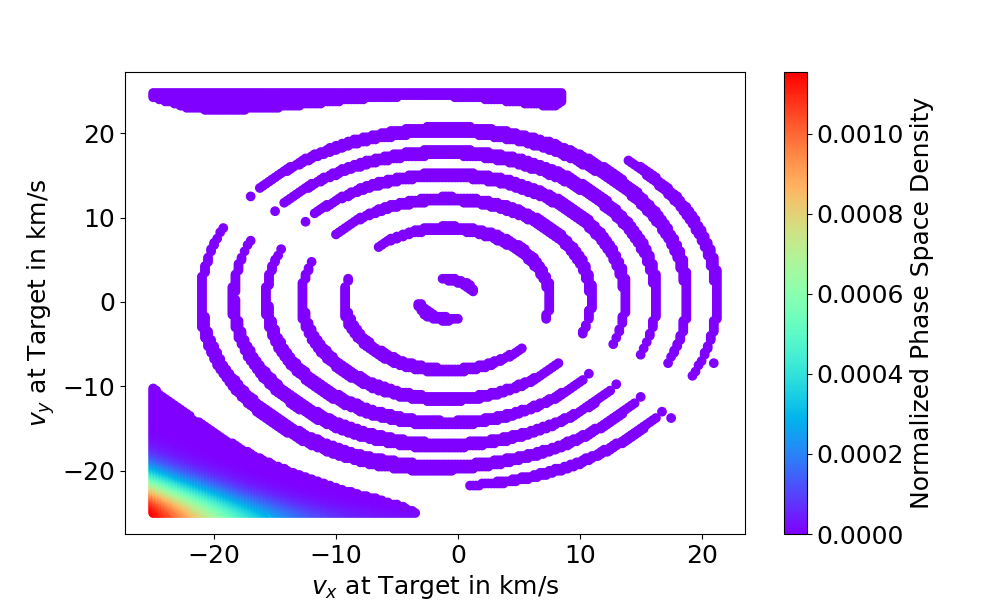}{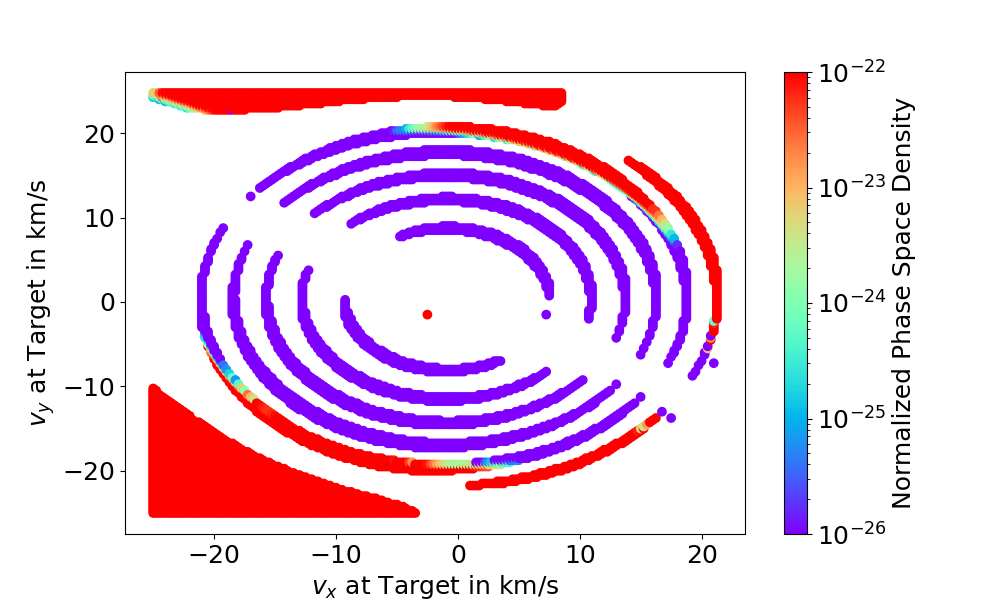}
    \caption{VDF plots (left: linear; right: log scale) at the same target location as Figure \ref{fig:center}, but at an earlier time $t = $-$1.5$ years (upper row), and a later time $t = 1.08$ years (lower row); the entire time range has an essentially flat $\mu = 0.66$ value due to the plateau near solar minimum of the value of $\mu$, as seen in Figure \ref{fig:rpfunct}. The plots show the evolution of the ring structures as described in the text.}
    \label{fig:earlylatecenter}
\end{figure}

\begin{figure}
    \plotone{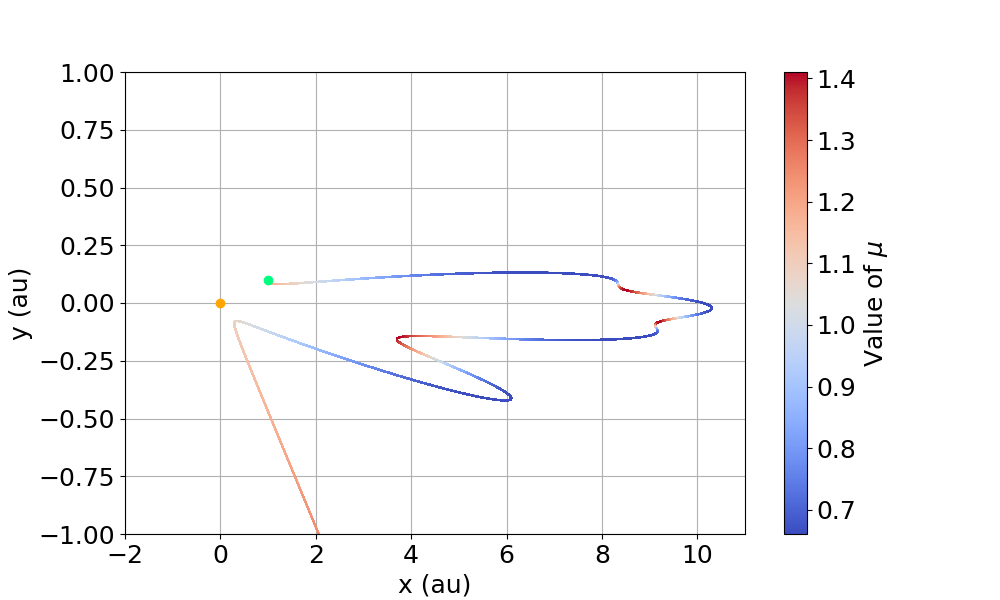}
    \caption{Example of an individual orbit that lingers in the region near the Sun for multiple solar cycles before intersecting the target point. The cycles can be counted by how many times the colors switch. The target point is located at $\mathbf{r} = (0.995 \text{, } 0.098)$ au, and the trajectory has a final velocity of $\mathbf{v}_0 = (0.9, 0.72)$ km s$^{-1}$ and reaches the target point at $t = 4.54$ years, when $\mu = 1.24$ ($f = 7.85 \times 10^{-61}$, perihelion $0.30$ au, age $60$ years).}
    \label{fig:longindorb}
\end{figure}

\begin{figure}[ht!]
    \plotone{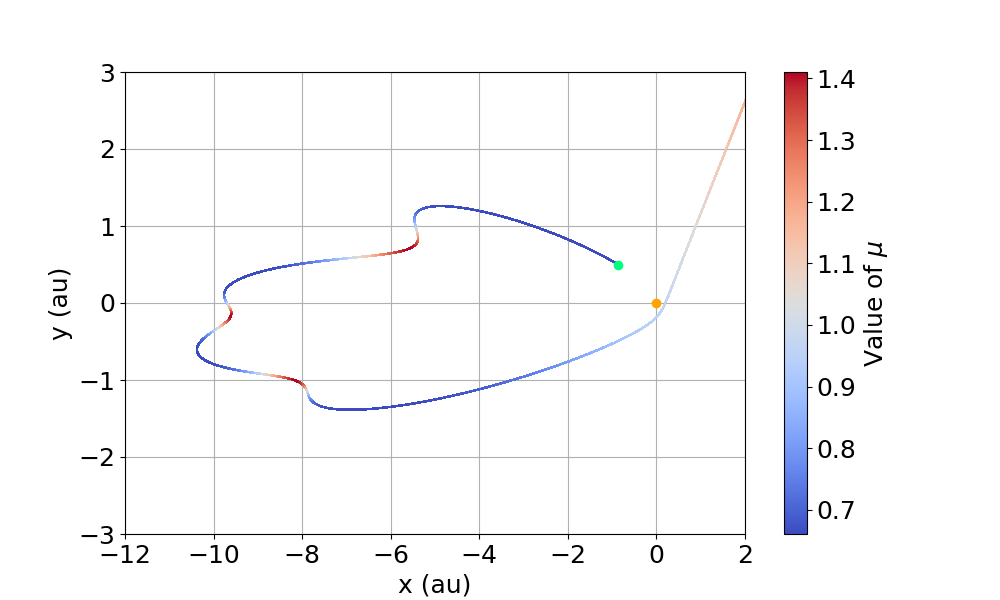}
    \caption{Individual trajectory sampled from the spatial and temporal point shown in the bottom row of Figure \ref{fig:earlylatecenter}, namely the point at $\mathbf{v}_0 = (21.0$, $-7.25)$ km s$^{-1}$ ($f = 4.04 \times 10^{-34}$, perihelion $0.29$ au, age $56$ years). This trajectory demonstrates an orbit that lingers in the region around the Sun for multiple cycles, like the one in Figure \ref{fig:longindorb}.}
    \label{fig:laterindorb}
\end{figure}

Investigating many target points and points in time, a pattern emerges that when the force from the Sun switches from repulsive to attractive, there begin to appear in phase space a series of concentric ring-like structures that correspond to the pseudo-bound orbits. These structures are the focus of Figures \ref{fig:center} - \ref{fig:earlylatecenter}.

These ring-like structures are associated with low velocities, and hence appear in a VDF region close to the origin, closer than the direct and indirect VDF features. They appear in VDF's at target points both upwind and downwind. For example, for the spatial and temporal point used in Figures \ref{fig:norpcomponents} and \ref{fig:componentsrp}, the center region with the radiation pressure force applied is plotted in Figure \ref{fig:center}. The same area in Figure \ref{fig:centerupwind} then provides some contrast, focusing on the low velocity region corresponding to the spatial target point given in Figure \ref{fig:dirindirupwind}. Note that the ring structures are interrupted in a coherent way so that a linear empty region forms whose corresponding velocities are close to radial velocities. We refer to this feature as “axis of exclusion.” The axis of exclusion appears in the area of these phase space plots where the pseudo-bound orbits lie (Figures \ref{fig:center}, \ref{fig:centerupwind}, \ref{fig:earlylatecenter}) because the associated trajectories are essentially linear and aimed at the Sun (zero angular momentum), with the Sun absorbing the atoms in those cases.

Figure \ref{fig:centerzoom} shows a zoomed in view of a portion of Figure \ref{fig:center}, so as to see the fine structure of the rings in higher resolution. In addition, Figure \ref{fig:earlylatecenter} shows the center of velocity space at the same spatial point but $1.6$ years earlier (upper row) than the time snapshot of Figure \ref{fig:center}, and $0.9$ years later (lower row). The ring structures first appear near the origin and expand outward in velocity space over time, such that the one nearly full ring structure in the top row of Figure \ref{fig:earlylatecenter} becomes the outermost ring in the other snapshots in Figures \ref{fig:center} and \ref{fig:earlylatecenter}. Subsequent rings appear near the origin and increase in radius, each ring corresponding to trajectories with an increased number of round trips around the Sun, with the outermost nearly full ring corresponding to one round trip.

Analyzing a variety of target points at 1 au, there are a few key properties that we observe about the ring structures when following them over time, largely revolving around their location and extent and how they accumulate through the attractive regime as time passes. Ring structures whose particles enter the region around the Sun and intersect with the target point within the same solar cycle adhere to these properties rigidly. However, it is possible for a trajectory to reach a local perihelion near the Sun first, be repelled, and reach the target point a few solar cycles later. These multi-cycle orbits don't yield as rigid of a VDF structure, and while they generally intersect the target point with greater magnitudes of velocity, this rule isn't universal. Examples of these multi-cycle orbits are given in Figures \ref{fig:longindorb} and \ref{fig:laterindorb}.

Ionization affects these ring structures preferentially. Trajectories corresponding to inner ring structures at later times are generally removed due to a large number of orbits often corresponding to tighter orbits and therefore particles being lost in the Sun. In addition, these inner ring structures are attenuated more strongly since the corresponding trajectories spend more of their time close to the Sun. Thus, the outer structures experience fewer losses than the inner structures, though all structures generally end up with lower PSD's than the cores of the direct component, and in many cases the indirect component as well. 

The genre of trajectories that execute an extended orbit in the region around the Sun after reaching a local minimum in radial distance are ionized depending on where they linger the most relative to the Sun. So, depending on the nature of the trajectory itself, this genre of pseudo-bound orbits is ionized in a way that doesn't adhere to any particular structure like the kind that only linger in the region around the Sun for around a cycle or less. They are, in general, ionized less since they spend less time in the region very close to the Sun (i.e., within $1$ au), but they are still generally ionized until the normalized PSD is on the order of $10^{-10}$ or less.

\begin{figure}[ht!]
    \plotone{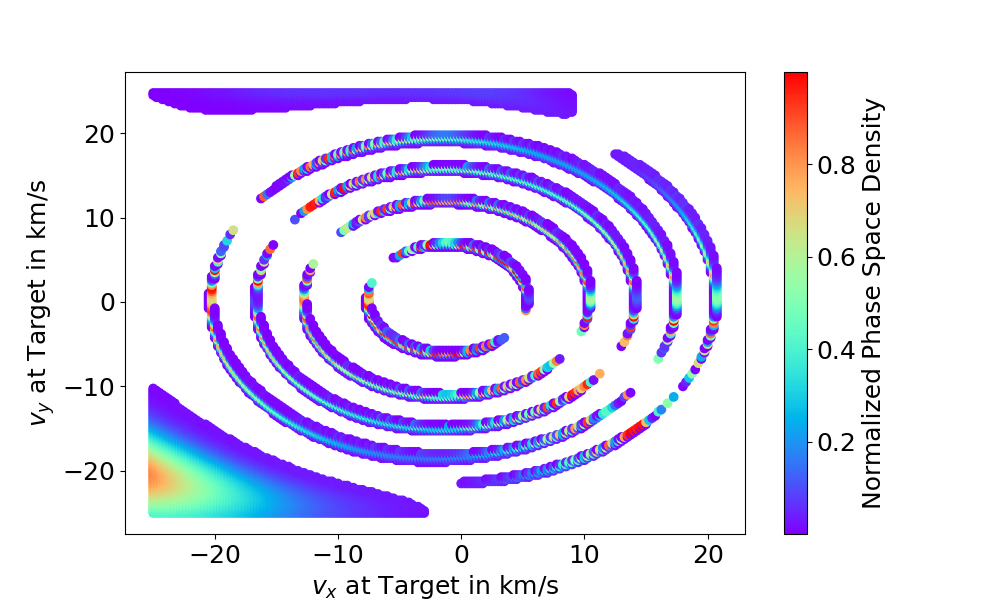}
    \caption{Same graph as Figure \ref{fig:center} with no ionization, which illustrates that the values of the phase space density associated with scattered points in the ring structure show no strong correlation.}
    \label{fig:centernopi}
\end{figure}

While the marked trend of outer ring points having higher PSD values appears with ionization present, the phase space density values without ionization associated with the points exhibit no consistent correlation. As seen in Figure \ref{fig:centernopi}, the only seemingly consistent trend is that there is a ridge of locally higher PSD values in the interior of the ring structures that is approximately symmetric with respect to radial distance from the velocity origin. However, where the arc of the ring approaches the axis of exclusion, all trends in the value of the normalized phase space density seem to disappear. 

In addition, it is worth remarking that the occurrence of pseudo-bound orbits, as well as the direct and indirect components in a time-dependent radiation pressure force, means that it is important to retain a relatively fine resolution in time points for the ionization integration process, as these trajectories can reach turning points in $r$ multiple times before reaching the target point, all of which can affect the integration process.

The ring structures, when present, have velocity magnitudes that are always less than those of the direct and indirect components at that instant in time, and thus correspond to lower energy particles. This conclusion also makes sense physically, as the phenomenon of pseudo-bound orbits arises from particles losing energy in the process of the force on them switching from repulsive to attractive, meaning a loss in energy as the potential energy goes from positive to negative non-adiabatically.

Additionally, due to ionization affecting these trajectories more heavily thanks to their proximity to the Sun, plenty of these trajectories would result in lower fluxes than their corresponding direct and indirect components. On average, the value of the ring structure PSD's differ from that of the direct component core PSD by a factor of at least $\sim 10^{-5}$ or more. This factor often corresponds to trajectories located on the first (outermost) ring structure - depending on the time and location at which we observe the target point, the factor differs. Most of the trajectories corresponding to phase space points in the inner ring structures are subject to extremely intense ionization before intersecting with the target point.

These two points bring up a potentially problematic issue with measuring particles following trajectories associated with pseudo-bound orbits in contrast to the usual set of direct and indirect trajectories. Not only are the pseudo-bound trajectories more scarce due to losses, they also exist within a lower energy range as the direct and indirect trajectories, such that existing spacecraft instruments may be unable to observe them. However, it is possible that particles occupying these pseudo-bound trajectories contribute to the background, and contribute to particle populations available for the generation of hydrogen through processes like charge exchange. Given that the ring structures cover a large range of angles from the origin in velocity space, we anticipate that after considering the shift in energy from moving to the non-inertial spacecraft frame, some of these trajectories may end up being measurable at explorer probes regardless. This notion, as well as the creation of neutral atoms from pseudo-bound orbits, warrant further study and more concrete calculations.

\begin{figure}[ht!]
    \plotone{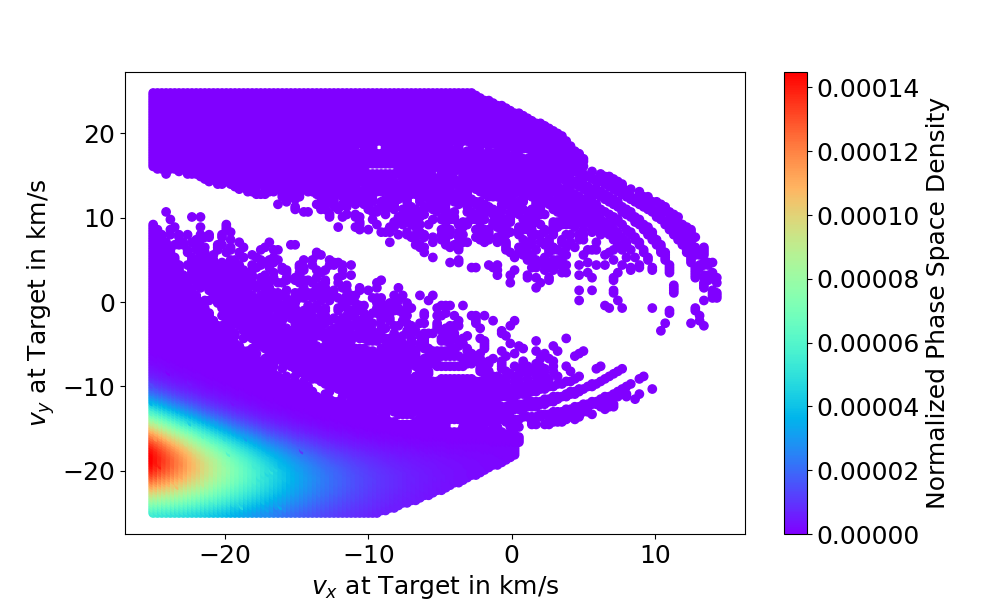}
    \plotone{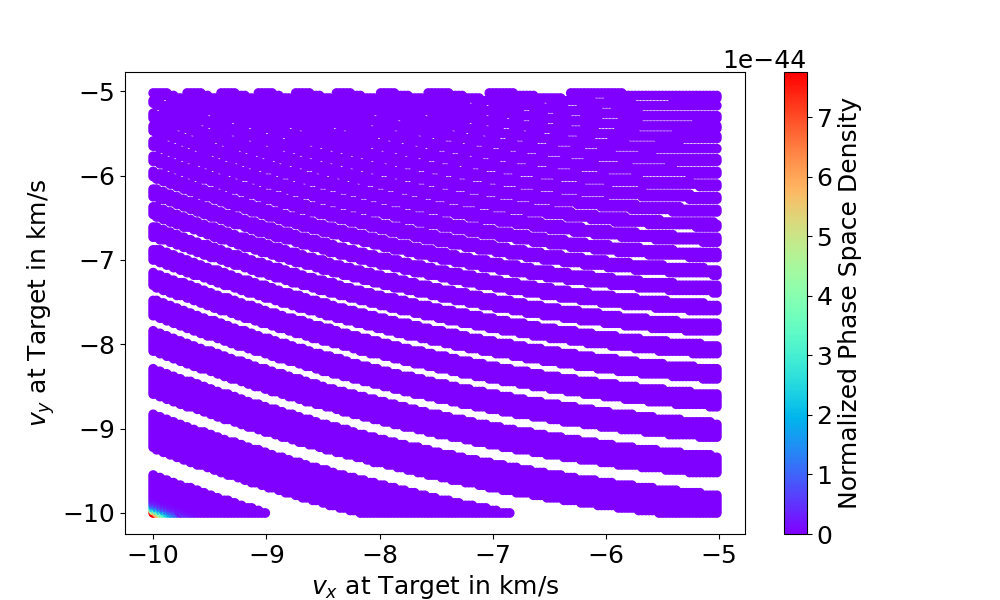}
    \caption{An example of the behavior of pseudo-bound orbit VDF structures at the time when the system is force free (at $t = 3.83$ years, when $\mu = 1.0$). The bottom plot is a more zoomed in version of the top plot.}
    \label{fig:forcefree}
\end{figure}

From a theoretical perspective, the existence of these ring-like structures is of particular interest, specifically how they manifest and the rules that they follow. They are surprisingly well-structured in phase space and stay coherent for an extended period of time throughout the cycle of $\mu (t)$. Even during later periods and into the time where the force becomes repulsive ($\mu > 1$), the structures persist, albeit tightly packed together in phase space to the point where high resolution in velocity is needed to distinguish them clearly, an effect that is presented in Figure \ref{fig:forcefree}. In the repulsive period, the structures shrink in the low velocity region and disappear.

\section{Conclusions} \label{sec:concl}

In this study, we analyze the effects of a time-dependent radiation pressure force on the trajectories of primary neutral hydrogen streaming into the heliosphere from the LISM. We focus on target points at 1 au distance from the Sun, and on trajectories that intersect at the target at the same instant in time. Analogous to the case of time-independent net force (see, e.g., \citet{mueller2012} for helium), we find that the so-called “direct” and “indirect” components have VDF peaks that shift both in magnitude and location in velocity space simply as a function of different spatial target points. Moreover, the time dependence of the net force on H atoms we treat in this paper in general adds additional VDF shifts and deformations, differently at each spatial point. The radiation pressure force, in the form presented here (Equation (\ref{musimple})), is oscillatory with the solar cycle and has a magnitude such that the force on particles has both periods of being net attractive and net repulsive. This net force and the time scale of oscillation, combined with the time scales of the trajectories in the region close to the Sun where the force is most influential, result in spatial target points where the direct and indirect components are no longer present during certain temporal regimes. This specifically happens when the force is or recently has been repulsive, as a repulsive force from the Sun forms an exclusion zone that largely disallows certain kinds of indirect orbits and direct orbits at far downwind target points.

In addition, we find that the non-adiabatic force transition can cause particles to lose energy in large amounts as they enter the region close to the Sun, resulting in orbits that are bound for one or multiple cycles of the radiation pressure force. These trajectories all intersect with their corresponding spatial target point starting after a certain period into the time regime when the force is attractive, and stop intersecting said points after the force has been repulsive for a period. These so-called pseudo-bound orbits arrange themselves into ring-like structures centered approximately around the origin in velocity space, with loose correlation between the location of the trajectory in these ring structures and the associated location of the trajectory in the Maxwellian at the injection plane. There are a number of behaviors associated with the appearance and arrangement of these ring structures that are consistent across all target points.

We also consider the effects of ionization on the trajectories, which involves integrating ionization rates for both charge exchange and photoionization along the trajectory to determine the attenuation of the normalized phase space density. This effect makes it such that pseudo-bound orbits that linger around the Sun for shorter periods (corresponding to more outer ring structures) are attenuated less, making a pattern among these structures. More importantly, we observe that the direct and indirect VDF components are attenuated to varying degrees depending on the spatial and temporal point at which they are observed. The direct components have their phase space densities reduced to less than $10\%$ of the original value, with the most attenuation generally happening during the repulsive regime for direct components that survive during that period. Meanwhile, indirect components have their phase space densities generally reduced by factors upward of $10^{5}$ compared to their original value, which is a result of indirect orbits broadly spending more time in the region around the Sun. The pseudo-bound orbits are attenuated a further few orders of magnitude or more in the value of the phase space density than even the indirects, at times much more, meaning they are less significant in terms of ability to be measured consistently.

This study offers a detailed look at the effects that a radiation pressure force periodic with the solar cycle alone has on the motion of primary neutral hydrogen through the heliosphere to points where explorer probes measure. While we do not propose to have created a comprehensive model, knowledge of these characteristic behaviors is important for identification and classification of more complex effects, which will occur for more comprehensive modeling. With awareness of the ring structures that appear in this environment and their characteristics, we will be able to better identify large and small scale structures of exotic orbits in a more realistic scenario.

One important future consideration is a transition to simulating in 3D, as both charge exchange (through the introduction of a non-uniform plasma distribution) and the radiation pressure force introduce asymmetry about the ISM-Sun axis that is suited for investigation in 3D. Another salient consideration is including the dependence on the radial velocity component $v_r$ due to Doppler shifting of solar radiation, corresponding to a non-flat solar Ly-$\alpha$ line profile. It would also be beneficial to conduct more data-driven simulations, using functions for $\mu$ and each $\beta$ that are modeled more precisely and accurately from experimental data so as to increase the accuracy of simulations relative to observations. These effects will be explored in a follow-up paper, which will also form a realistic basis to compare with bulk parameter modeling results by \citet{izmodenov2013}, and with IBEX measurements.

The results of this study show promise for analysis of trajectories of primary neutral hydrogen as can be potentially observed by current and future spacecraft instrumentation. We can characterize the approximate location in velocity space of the direct and indirect components as well as other structures that arise from pseudo-bound orbits, all according to a model that roughly approximates experimental observations of the time-dependent radiation pressure force that oscillates with the solar cycle. Use of the trajectory method will allow us to accurately assess the location in phase space, and thus the energy, of neutral hydrogen at $1$ au, which will allow for greater accuracy in predicting measurements made by explorer probes. Moving toward greater and more precise simulation parameters with an optimized code, as well as consideration of geometry and viewing angles of detectors in their spacecraft frame, will allow for more accurate assessments and correlation to measurements.

\begin{acknowledgements}
The authors gratefully acknowledge support for this research by NASA grant 80NSSC21K1681. This research was supported by the International Space Science Institute (ISSI) in Bern, through ISSI International Team project \#541 (Distribution of Interstellar Neutral Hydrogen in the Sun's Neighborhood).
\end{acknowledgements}

\software{NumPy \citep{numpy2020}
    MatPlotLib \citep{matplotlib2007}
    SciPy \citep{scipy2020}
    tqdm \citep{tqdm2021}}

\bibliography{main}

\end{document}